\documentclass[%
  reprint,
  amsmath,amssymb,
  aps,
  prc,
  nofootinbib,
  floatfix,
 superscriptaddress, longbibliography
]{revtex4-2}
\usepackage[utf8]{inputenc}
\usepackage{graphicx}
\usepackage{dcolumn}
\usepackage{bm}
\usepackage{xspace}
\usepackage{tcolorbox}
\usepackage{tabularx}
\usepackage{enumitem}

\usepackage{comment}

\newcommand{\gev}{\ifmmode \xspace
\ensuremath{\textnormal{GeV}\xspace} \else GeV \fi}

\usepackage[colorlinks=true, linkcolor=blue, citecolor=blue, urlcolor=blue]{hyperref}

\begin{document}

\title{Physics Opportunities with a Fixed-Target Program at the Electron-Ion Collider}

\author{C.-J. Naïm}
\email{Corresponding author: charlesjoseph.naim@stonybrook.edu}
\affiliation{Center for Frontiers in Nuclear Science (CFNS), Department of Physics and Astronomy, Stony Brook University, Stony Brook, NY 11794, USA}

\author{A. Sorensen}
\affiliation{Facility for Rare Isotope Beams, Michigan State University, East Lansing, MI 48824, USA}

\author{D. Brown}
\affiliation{National Nuclear Data Center, Brookhaven National Laboratory, Upton, NY 11973, USA}

\author{D. Cebra}
\affiliation{Department of Physics and Astronomy, University of California, Davis, Davis, CA 95616, USA}

\author{R. Corliss}
\affiliation{Center for Frontiers in Nuclear Science (CFNS), Department of Physics and Astronomy, Stony Brook University, Stony Brook, NY 11794, USA}

\author{J. M. Durham}
\affiliation{Los Alamos National Laboratory, Physics Division, Los Alamos, NM 87545, USA}

\author{R. Vogt}
\affiliation{Nuclear and Chemical Sciences Division, Lawrence Livermore National Laboratory, Livermore, CA 94551, USA}
\affiliation{Department of Physics and Astronomy, University of California, Davis, Davis, CA 95616, USA}

\begin{abstract}
A fixed-target program at the Electron-Ion Collider (EIC) would broaden the facility's scientific reach by providing key measurements for studies of cold nuclear matter (CNM), the QCD phase diagram, and nuclear reactions relevant for space radiation. Constraining CNM effects is essential for interpreting observables in proton-nucleus ($p+A$) and nucleus-nucleus ($A+A$) collisions, yet these effects are poorly understood at low center-of-mass energies. In particular, the range $\sqrt{s}\approx10$–$20~\gev$, where several CNM effects may compete at similar scales, has not been explored with high statistical precision. Mapping the QCD phase diagram similarly requires high-statistics $A+A$ data with broad rapidity and transverse-momentum coverage to probe the onset of deconfinement and the possible location of the QCD critical point (CP). Currently, such data are limited at $4.5~\gev < \sqrt{s_{NN}} < 7.7~\gev$. A fixed-target program at the EIC would fill these gaps, providing CNM baselines and complementary data for QCD CP studies. By delivering high luminosity and systematic measurements across a broad range of nuclear targets, the program would link $p+A$ and $A+A$ systems at the same center-of-mass energies, enabling a unified, quantitative description of cold QCD matter and clarifying the interpretation of QGP signatures in $A+A$ collisions at low energies where comparable $p+A$ data are lacking. Finally, the program would offer a unique opportunity to measure nuclear cross sections critical for improving cosmic-ray models, including studies of space-radiation protection for both autonomous spacecraft and long-duration human spaceflight.
\end{abstract}

\maketitle


\section{Introduction}
\label{sec:introduction}

The Electron–Ion Collider (EIC) will be the next-generation collider facility in the United States, designed to probe the internal structure of hadronic matter with unprecedented resolution with electron–nucleus~($e+A$) collisions~\cite{AbdulKhalek:2021gbh}. Operating at center-of-mass energies between 20 and 140~GeV, the EIC will enable precision studies of how nucleon mass and spin emerge from QCD interactions among quarks and gluons. Its versatile hadron-beam program will enable systematic studies of the dependence of strongly interacting phenomena on nuclear mass~$A$. The availability of polarized proton, deuteron, and light-ion beams will significantly extend the physics reach to nuclear spin-dependent observables~\cite{Atoian:2025dib}. Together, these capabilities establish the EIC as a comprehensive facility for exploring the internal structure of nucleons and nuclei across a wide kinematic range.

Beyond its primary role as a high-luminosity~$e+A$ collider, the EIC offers unique opportunities to extend studies of strong interactions in hadron–nucleus~($p+A$) and nucleus–nucleus~($A+A$) collisions by operating in a fixed-target configuration. Such a program, using proton and ion beams on stationary targets, would provide a complementary means of investigating effects arising both from properties of cold nuclear matter (CNM) and from the dynamics of a strongly interacting, hot, and possibly deconfined medium.

The study of CNM effects has a long experimental history, beginning with fixed-target measurements in the late 1970s~\cite{Branson:1977ci, Antipov:1977ss, Anderson:1979tt} establishing that the nuclear environment can substantially modify hard processes. A landmark result was obtained by the NA3 experiment at CERN~\cite{Badier:1983dg}, which observed a strong suppression of $J/\psi$ production in pion- and proton–nucleus collisions, demonstrating that heavy quark–antiquark pair formation and propagation are sensitive to the nuclear environment. These measurements established quarkonia as key probes of CNM effects.

More recently, the SeaQuest (E906) experiment at Fermilab~\cite{Lin:2017eoc}, operating at a center-of-mass energy of $\sqrt{s_{NN}}=15$~GeV, reported indications of a suppression of the Drell–Yan yield in heavy nuclei at large Feynman-$x$, $x_F$. If confirmed, this observation would constitute the first direct experimental evidence for sizable initial-state parton energy loss in CNM at low energies~\cite{Arleo:2018zjw}. Despite these advances, the global understanding of CNM effects remains incomplete, as the relative importance of different mechanisms and their energy dependence are still poorly constrained.

A recent review of CNM effects~\cite{Arleo:2025oos} emphasized the need for a coherent experimental program capable of isolating CNM effects across multiple hard probes, including Drell–Yan, open heavy flavor, and quarkonium production. In particular, precise proton–nucleus measurements at $\sqrt{s_{NN}}\lesssim20$~GeV, where CNM effects are expected to play a dominant role, are largely absent. Although some $p+A$ data exist at similar energies~\cite{Badier:1983dg, Lin:2017eoc, NA60:2010wey}, the available measurements remain sparse and limited. In particular, they lack a controlled study of the nuclear mass dependence and do not provide continuous beam-energy coverage. Consequently, CNM effects on hadronic observables remain largely unexplored at these energies. This lack of key measurements limits the interpretation of nuclear modification of hard probes observed in nucleus–nucleus collisions at the Relativistic Heavy Ion Collider (RHIC)~\cite{Adare:2012qf, PHENIX:2013pmn, PHENIX:2019brm, PHENIX:2022nrm, STAR:2013kwk, STAR:2021zvb} and at the Large Hadron Collider (LHC)~\cite{Adam:2015jsa, ALICE:2014cgk, ALICE:2014ict, ALICE:2022zig, ALICE:2020vjy, ALICE:2019qie, LHCb:2013gmv, LHCb:2016vqr, LHCb:2017ygo, LHCb:2024taa, LHCb:2018psc, LHCb:2014rku, CMS:2017exb, CMS:2018gbb, CMS:2018bbk, CMS:2022wfi, Jonas:2026yoz}, where effects arising from CNM and from a deconfined quark–gluon plasma (QGP) are intrinsically entangled.

Beyond hard probes, inclusive hadron spectra in~$A+A$ collisions are also expected to be modified by CNM effects~\cite{STAR:2017sal, STAR:2019bjj, STAR:2026kqj}, highlighting the importance of systematic $p+A$ measurements to understanding the dynamics of hot nuclear matter. At low beam energies, $\sqrt{s_{NN}}\lesssim20$~GeV, the coherence length of partonic interactions becomes comparable to the nuclear size and several CNM mechanisms, such as modifications of the nuclear parton distribution functions (nPDFs), parton energy loss, and nuclear absorption, may all substantially contribute to the observed nuclear modification~\cite{Arleo:2025oos}. A quantitative description of these effects is essential for developing a consistent picture of parton production, propagation, and final-state hadron formation over the entire kinematic range.

The critical domain of missing measurements corresponds to center-of-mass energies~$\sqrt{s_{NN}} \approx 10$--$20$~GeV. This range overlaps with the reach of a proposed fixed-target program at the EIC, which offers an opportunity to close this gap with precise and systematic studies of CNM effects over a broad range of nuclear targets and beam energies. The program could directly compare $e+A$ in collider mode and $p+A$ in fixed-target mode with the same detector. Despite the difference in $\sqrt{s}$, the nucleus momentum fraction~$x$ and the hard scale~$Q^2$ can be comparable. The shared systematics would enable direct tests of the universality of CNM effects in lepton- and hadron-induced processes.

In addition, a fixed-target program at the EIC would also enable $A+A$ collisions at energies comparable to $p+A$, providing direct access to the transition from hadron- to parton-level interactions in small systems and from cold nuclei to hot and dense QCD matter in larger systems. 
While a partonic interpretation can be ruled out at the lowest energies, contributions from parton interactions are expected to play an increasingly important role as $\sqrt{s_{NN}}$ and the system size increase. Complementary information on nuclear matter effects in this energy range has been obtained by the STAR experiment at RHIC through the Beam Energy Scan~(BES) and Fixed Target~(FXT) programs~\cite{STAR:2017sal, STAR:2020dav, STAR:2021yiu}, which explored Au+Au collisions over $\sqrt{s_{NN}} = 3$--$39~\gev$. In this regime, the system evolves from a hadron-dominated medium toward QGP. However, the onset and properties of hot deconfined QCD matter at these energies remain largely unknown, and the role of CNM effects in both the confined and deconfined phases is not clearly understood since no corresponding $p+A$ data have been collected. As a result, this energy range is a critical domain for disentangling CNM effects from hot-medium phenomena, affecting studies of the QCD equation of state~(EOS), the search for the QCD critical point~(CP), and studies of QGP properties such as transport coefficients.

Moreover, the kinematic limitations of the STAR detector in the fixed-target mode severely reduce sensitivity in the region $4.5~\gev < \sqrt{s_{NN}} < 7.7~\gev$, precluding meaningful coverage of a domain that is of particular interest for the QCD CP. Indeed, recent advances in theoretical predictions for the QCD CP suggest the importance of the region of the QCD phase diagram explored in that collisions energy range~\cite{Mukherjee:2019eou,Fu:2019hdw,Gao:2020fbl,Schmidt:2022ogw,Zambello:2023ptp,Basar:2023nkp,Clarke:2024ugt}. Moreover, multiple phenomenological approaches have also recently pointed to the same region~\cite{Hippert:2023bel,Shah:2024img,Sorensen:2024mry}.

Among existing or planned facilities, a fixed-target program at the EIC could provide measurements that smoothly connect the RHIC BES fixed-target and collider-mode results, enabling a definitive statement about the structure of the QCD phase diagram in regions accessible by terrestrial experiments. Current and future programs such as NA60+ at the CERN SPS~\cite{NA60:2022sze} and the CBM experiment at FAIR~\cite{Agarwal:2023otg} aim to explore complementary regions of the phase diagram. In particular, CBM will probe dense baryonic matter at lower collision energies, $\sqrt{s_{NN}} \approx 2.9$--$4.9$ GeV, with dedicated measurements targeting the search for the QCD phase transition and the CP, while NA60+ will cover the intermediate range $\sqrt{s_{NN}} \approx 6$–$17$ GeV using a dedicated muon spectrometer.
Nevertheless, a fixed-target program at the EIC would be the only one covering this range with one detector.

Overall, a unified experimental description across $e+A$, $p+A$, and $A+A$ systems is essential to test the universality of CNM effects and to map the QCD phase diagram. A fixed-target program at the EIC would provide key data spanning small to large systems and low to moderate energies: $\sqrt{s} \sim 7$--$23~\gev$ for $p+A$ and $\sqrt{s_{NN}} \sim 3$--$14~\gev$ for $A+A$ collisions. This regime is key to understanding both cold and hot QCD matter. The broad range of collision systems and energies available at the EIC would make it possible to study the transition from hadron-level to parton-level dynamics in small systems and from cold nuclei to hot and dense QCD matter in larger systems.

In addition to its impact on fundamental QCD studies, a fixed-target program at the EIC would also provide valuable data for applications beyond collider physics. In particular, precise measurements of nuclear interaction cross sections at low beam energies are directly relevant for modeling galactic cosmic-ray interactions with matter~\cite{Maurin:2025gsz}. Such interactions govern the production of secondary particles in shielding materials and spacecraft structures and are critical inputs for radiation transport calculations used in space-radiation protection studies. By enabling systematic measurements with light and intermediate-mass nuclei over a broad energy range, a fixed-target program at the EIC would therefore contribute not only to a unified understanding of cold and hot QCD matter, but also to improved predictive capabilities for radiation shielding for both autonomous and long-duration human spaceflight.

Given the breadth of the research opportunities outlined above, a fixed-target program at the EIC would significantly broaden the facility's scientific reach and maximize its potential for groundbreaking measurements.
By concentrating on $p+A$ and $A+A$ collisions in the energy range where multiple mechanisms contribute simultaneously, such a program would bring together communities that have so far largely pursued separate lines of investigation.
In particular, CNM effects are directly relevant both for the interpretation of hard probes and for observables associated with the QCD critical point.
A fixed-target program would therefore strengthen connections across subfields and reinforce the role of the EIC as a central facility for nuclear physics in the United States and worldwide.   
\section{Physics Case}
\label{sec:section1}

In this section, we highlight the physics opportunities enabled by a fixed-target program at the EIC. Rather than providing a comprehensive survey, we focus on measurements that are either unique to or significantly enhanced by a fixed-target configuration, and highlight their impact within the broader EIC physics program.

\subsection{Cold Nuclear Matter}

A fixed-target program at the EIC would provide a unique opportunity to systematically investigate CNM effects as a function of projectile energy and nuclear size~\cite{Arleo:2025oos}. By varying the beam energy and target species, such measurements would directly probe the energy dependence of CNM effects in a regime where parton energy loss~\cite{Neufeld:2010dz, Arleo:2012hn, Arleo:2018zjw}, transverse-momentum broadening~\cite{Johnson:2000dm, Arleo:2020rbm}, nuclear absorption~\cite{Lourenco:2008sk, Vogt:2011zzb}, and intrinsic charm~\cite{Vogt:2023plx, Vogt:2021vsc} are all expected to to be important.

\begin{figure} \centering \includegraphics[width=1\linewidth]{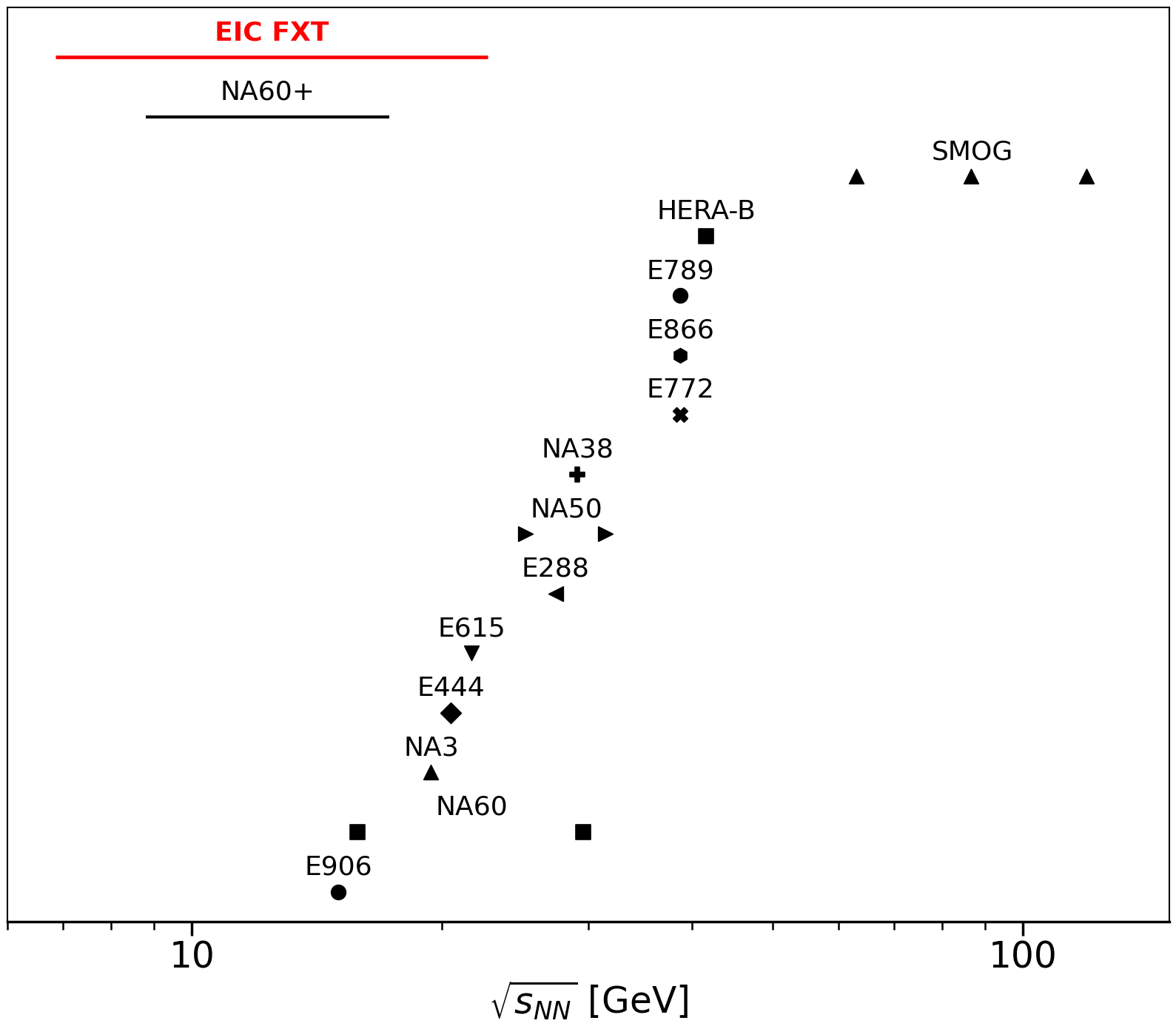} \caption{ The $\sqrt{s_{NN}}$ coverage of $p+A$ collisions in fixed-target experiments: NA3~\cite{Badier:1983dg}, NA38~\cite{NA38:1998lyg}, NA50~\cite{NA50:2006rdp, NA50:2003fvu}, NA60~\cite{NA60:2010wey}, E444~\cite{Anderson:1979tt}, E772~\cite{Alde:1990wa, Alde:1991sw}, E789~\cite{Kowitt:1993ns, Leitch:1995yc}, E866~\cite{Leitch:1999ea}, E906 (SeaQuest)~\cite{Lin:2017eoc}, HERA-B~\cite{HERA-B:2006bhy, HERA-B:2008ymp}, SMOG (LHCb)~\cite{LHCb:2022sxs, LHCb:2018jry}, NA60+~\cite{NA60:2022sze}. The projected energy coverage of the proposed EIC fixed-target measurements is shown in red.} \label{fig:energy_pA} \end{figure}

Measurements accessible in fixed-target $p+A$ collisions at the EIC (see Fig.~\ref{fig:energy_pA}) are essential for establishing baselines for quarkonium and open heavy-flavor and light-meson observables and for disentangling CNM effects from hot-medium phenomena observed in $A+A$ collisions. At low~$\sqrt{s}$, the hadron formation time becomes comparable to the nuclear size, enhancing sensitivity to nuclear absorption and hadronization effects inside the nucleus. Measurements in this energy regime also probe nPDFs at large momentum fractions~$x$, a region that remains poorly constrained, especially for gluons~\cite{Eskola:2016oht,Eskola:2021nhw,deFlorian:2011fp,AbdulKhalek:2019mzd,Kovarik:2015cma}. Performing systematic measurements over a broad energy range would further discriminate between competing quarkonium production models which differ most strongly at lower energies~\cite{Vogt:2001ky,Feng:2015cba}.

\begin{figure}
\centering \includegraphics[width=1\linewidth]{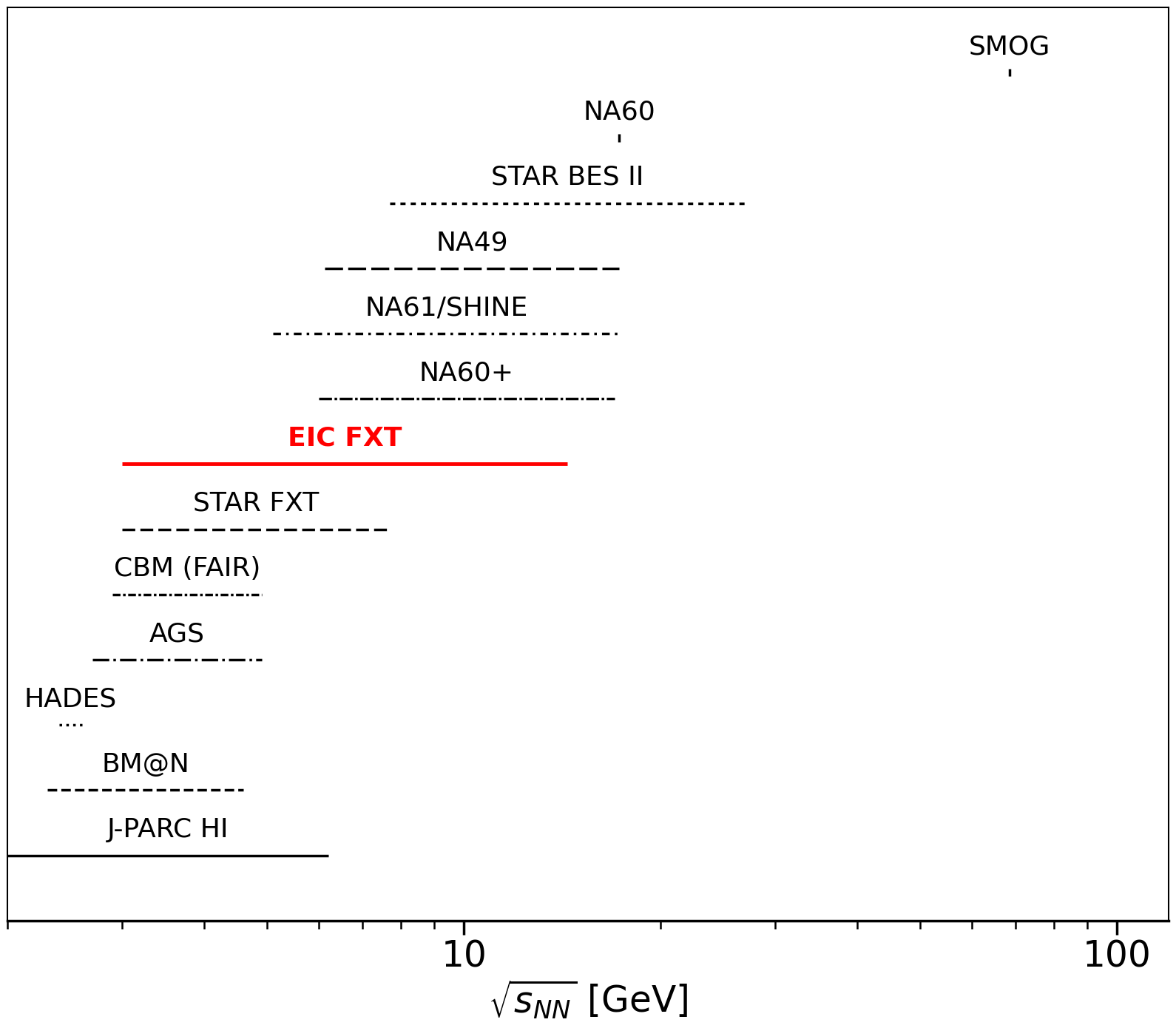} 
\caption{ The $\sqrt{s_{NN}}$ coverage in fixed-target $A+A$ collisions: HADES~\cite{HADES:2020ver}, BM@N~\cite{Baranov:2018cdk}, AGS~\cite{Odyniec:2013aaa}, STAR FXT~\cite{STAR:2020dav,STAR:2021yiu, STAR:2024zvj, STAR:2024znc, STAR:2022etb, STAR:2021fge}, STAR BES II~\cite{STAR:2017sal,STAR:2013gus}, NA49~\cite{NA49:2006gaj}, NA61/SHINE, NA60~\cite{NA60:2009una}, NA60+~\cite{NA60:2022sze}, CBM~\cite{Agarwal:2023otg}, J-PARC heavy-ion program~\cite{Koch:1986ud,Ozawa:2022sam}, SMOG~\cite{LHCb:2022qvj, LHCb:2025ixz}. The projected energy coverage of the proposed EIC fixed-target measurements is shown in red. 
}
\label{fig:AA_energies} 
\end{figure}

A fixed-target configuration at the EIC would also enable $A+A$ collisions at $\sqrt{s_{NN}} \sim 3$--$14~\gev$, covering the high–baryon-density region of the QCD phase diagram that has been explored by multiple experiments worldwide (see Fig.~\ref{fig:AA_energies}), including the RHIC BES collider and fixed-target measurements where the analysis is still ongoing.
This energy range corresponds to moderate temperatures and large net-baryon densities, where the nature of the QCD phase transition remains an open question.

Existing measurements at RHIC and the CERN SPS have revealed strong energy and system-size dependencies of quarkonium suppression, open heavy-flavor production, and low-mass dilepton yields~\cite{STAR:2023nos, STAR:2018xaj, NA60:2006ymb, NA60:2007lzy}. These observables are sensitive to color screening, regeneration, in-medium spectral modifications, and CNM effects~\cite{Vogt:2005ia, Zhao:2010nk}. However, their interpretation remains limited by the absence of precise systematic baselines at comparable energies, in particular of initial-state effects, formation-time dynamics, and hadronization in a high-baryon-density environment.

A combination of $p+A$ and $A+A$ fixed-target collisions at the EIC would address current limitations in separating CNM effects from genuine hot-medium contributions. High luminosities and modern detectors would allow precise measurements of light mesons, dilepton spectra, and flow observables in $A+A$ collisions. Performing measurements in $p+A$ and $A+A$ collisions at similar energies within the same detector setup would provide a well-controlled reference for CNM effects. This approach would reduce experimental systematic uncertainties and allow a clear separation of CNM contributions from strongly interacting matter in larger systems. 

\subsection{QCD Phase Diagram}

The recently completed RHIC BES and FXT programs enabled measurements of ${\rm Au}+{\rm Au}$ collisions for $\sqrt{s_{NN}} = 7.7$--$39~\gev$ in collider mode~\cite{STAR:2017sal,STAR:2013gus} and for $\sqrt{s_{NN}} = 3.0$--$7.7~\gev$ in fixed-target mode~\cite{STAR:2020dav,STAR:2021yiu, STAR:2024zvj, STAR:2024znc, STAR:2022etb, STAR:2021fge}.
The central goal of these experiments was to locate the phase boundary between hadronic and partonic matter and determine whether these two phases of QCD matter can coexist or, equivalently, whether there is a QCD critical point~\cite{STAR:2020dav, STAR:2017sal, STAR:2020tga, STAR:2026kqj, Bzdak:2019pkr, Stephanov:2004wx, Sorensen:2023zkk, Du:2024wjm, Chen:2024aom}.
Key observables include higher-order cumulants of conserved charges, collective flow coefficients, and electromagnetic probes such as low-mass dileptons.

A fixed-target program at the EIC would further support these studies in several ways.
First, an important avenue that has not been explored in the BES program is $p+A$ collisions, which would provide important baselines for $A+A$ measurements. For example, small-system studies of observables such as triangular flow~\cite{Abdulhamid:2024uky}, whose origin has been linked to sub-nucleonic fluctuations and early-time dynamics~\cite{Zhao:2023rss}, could help disentangle genuine collective effects from other correlations. In particular, BES-I measurements showing that triangular flow remains finite even in low-energy $A+A$ systems and only disappears in peripheral collisions~\cite{STAR:2016vqt} raise the question whether it reflects medium response or initial-state correlations. A systematic comparison to $p+A$ collisions at the same energies would clarify this issue and help separate CNM contributions from genuine signatures of a strongly interacting medium. Similarly, a comparison to a $p+A$ baseline could help interpret cumulant measurements~\cite{STAR:2023zhl, STAR:2025rpo, ALICE:2026ajq}, which are one of the primary observables for locating the QCD CP on the QCD phase diagram.

The fixed-target program at the EIC would effectively bridge the energy gap between RHIC FXT data and the collider-mode BES data. Although not originally designed for fixed-target operation, STAR has successfully collected high-statistics data for $\sqrt{s_{NN}} = 3.0$--$7.7~\gev$. However, due to detector acceptance and fixed-target geometry, full (symmetric) midrapidity coverage is only achieved at $\sqrt{s_{NN}} = 3.0~\gev$~\cite{Almaalol:2022xwv}. This limitation significantly impacts key observables for the QCD critical point search, particularly net-proton cumulants, where the acceptance dependence carries important information. This highlights the need to extend these studies with higher statistics and to explore this kinematic region with improved acceptance. 

The EIC fixed-target energy range would therefore bridge the gap between the highest STAR FXT data and the collider-mode. In particular, measurements at $4.5~\gev < \sqrt{s_{NN}} < 7.7~\gev$ would provide continuity across the domain where theoretical predictions for the QCD critical point are concentrated in recent state-of-the-art theoretical analyses~\cite{Mukherjee:2019eou,Fu:2019hdw,Gao:2020fbl,Schmidt:2022ogw,Zambello:2023ptp,Basar:2023nkp,Clarke:2024ugt,Hippert:2023bel,Shah:2024img,Sorensen:2024mry}. Together with improved kinematic coverage over the entire BES range, this would substantially strengthen the experimental foundation for mapping the QCD phase structure.

Moreover, a unique aspect of an EIC fixed-target heavy-ion program is the availability of polarized light-ion beams, including deuteron, helium, and lithium~\cite{Atoian:2025dib}. Collisions with polarized nuclei would provide the first measurements of spin-dependent effects in dense QCD matter such as spin–orbit interactions, polarization transport, and spin-dependent modifications of parton energy loss and hadronization. Polarization-sensitive observables, including spin asymmetries, vector-meson spin alignment, and spin-dependent flow, are qualitatively new probes of the microscopic dynamics of baryon-rich QCD matter that are inaccessible at existing heavy-ion facilities and would open a new dimension in the study of dense QCD matter.

Together with baselines from $p+A$ collisions, the $A+A$ collisions in the EIC fixed-target energy range would thus significantly advance our understanding of the behavior of dense nuclear matter, the emergence of deconfined matter, and the QCD phase structure, complementary to the RHIC physics programs.

\subsection{Space Radiation Studies}

Recent studies have emphasized the need for precise nuclear cross-section measurements to improve modeling of galactic cosmic rays. Such models are important for predicting radiation levels encountered by astronauts and spacecraft during long-duration human spaceflight and for devising mitigation strategies~\cite{Chancellor03042021, 10.1371/journal.pbio.3000669, Smith:2023fspas, Maurin:2025gsz}.

Interactions of nuclei dominant in cosmic rays (H, He, C, O, Ne, Mg, Si, and Fe; see Fig.~2 in Ref.~\cite{Chancellor03042021} and Fig.~2 in Ref.~\cite{Maurin:2025gsz}) with light and intermediate nuclear targets produce secondary particle showers containing light fragments, neutrons, and pions over a projectile energy range from approximately 0.1 to 50 GeV per nucleon~\cite{NORBURY2012315}. These secondaries play a central role in estimates of radiation dose behind shielding materials.

Interestingly, although the flux of heavier nuclei is orders of magnitude lower than that of lighter particles, their secondary showers are sufficiently large that they dominate shielding considerations. Existing data are sparse and often lack the precision required (typically better than 20\%) for reliable radiation transport simulations. Data are needed for nuclei in structural and shielding components (Fe, Ni, Cr, and Al), electronics (Si, Ge, Ga, N), and the human body (C, O, N, Ca). While compilations of such data exist~\cite{luoni2021totalnuclearreactioncrosssection}, global coverage remains insufficient for accurate shielding modeling~\cite{NORBURY2012315}.

In this context, fixed-target $A+A$ measurements at the EIC, especially at forward rapidity, would provide essential experimental constraints. Systematic measurements of inclusive and double-differential pion, neutron, and light-ion production cross sections, using projectiles such as He, O, Si, and Fe on targets representative of spacecraft materials, would substantially reduce current uncertainties. Such data would directly enhance the predictive power of radiation transport codes and support planning and risk assessment for future autonomous and crewed space missions.
\section{Energy and luminosity}
\label{sec:section2}

The EIC is designed to provide flexibility in beam species, energy, and polarization. In its baseline configuration, the hadron beam energy can be varied from about $25$ to $275$~GeV for protons and from about $3.85$ to $110$~GeV per nucleon for gold nuclei~\cite{RHICRunOverview}.

In the fixed-target configuration considered here, the hadron beam would impinge on a thin nuclear target installed along the hadron-beam direction. The accessible $p+A$ energy range is $\sqrt{s_{NN}} \simeq 7$–23~GeV, while the corresponding $A+A$ range is $\sqrt{s_{NN}} \simeq 3$--14~GeV, depending on the projectile species and beam energy. Various ions can be used as projectiles, from light nuclei up to uranium.

These energy domains overlap with the region where studies of CNM effects and of the properties of hot and dense QCD matter are needed, and where existing experimental constraints remain limited. 
In addition to high luminosity, a key and unique feature of the EIC is the availability of highly polarized hadron beams, including protons, deuterons, and light nuclei such as $^3$He and lithium isotopes, with $\sim70\%$ polarization expected in the baseline collider program~\cite{Atoian:2025dib}.

Therefore, a fixed-target mode at the EIC would also open the door to a new experimental program using polarized proton and light-ion beams on a nuclear target. Spin-dependent measurements in these collisions would access a largely unexplored kinematic regime, enabling studies of spin–orbit effects, spin-dependent nPDF modifications, and transport of polarization at moderate~$\sqrt{s_{NN}}$.

\begin{figure}
\centering
\includegraphics[width=1\linewidth]{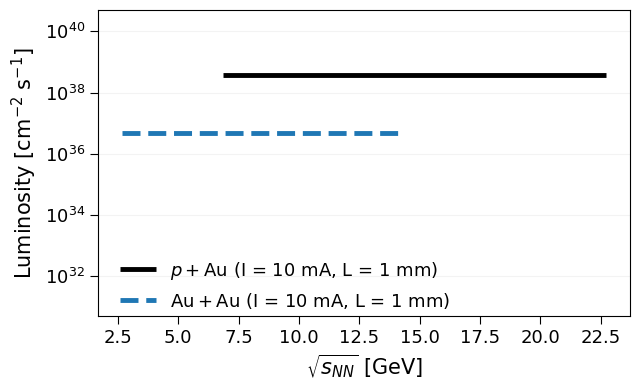}
\caption{
Nucleon–nucleon center-of-mass energy $\sqrt{s_{NN}}$ and estimated instantaneous fixed-target luminosities for representative $p+\mathrm{Au}$ and $\mathrm{Au}+\mathrm{Au}$ collisions at the EIC, assuming a 1~mm thick gold foil and an effective beam current of $I_{\mathrm{eff}} = 10$~mA.
}
\label{fig:luminosity_fixed_target}
\end{figure}

The STAR collaboration successfully operated a fixed-target program with RHIC beams using a $\sim 1$ mm thick gold foil as a target~\cite{Liu:2021ypa}. That program required dedicated beam steering onto the target, separate from the RHIC collider mode. Assuming a target of the same thickness at the EIC, we estimate the luminosity as
\begin{equation}
\mathcal{L} = \Phi n_{\mathrm{Au}}
\end{equation}
where $\Phi = I/(Z e)$ is the incident beam flux, $I$ is the circulating beam current, $Z$ the ion charge state, and $n_{\mathrm{Au}}$ the target areal density, which depends linearly on the target length $L$. At the EIC, the nominal circulating beam current is $I_p = 1$~A for protons and $I_{\mathrm{Au}} = 0.57$~A for gold ions in collider operation~\cite{EIC_ParameterList_2026}. In realistic solid fixed-target running, only a small fraction of the circulating beam would be intercepted by the target in order to preserve collider performance and control detector occupancies. Assuming effective beam currents at the percent level, corresponding to $I_{\mathrm{eff}} \sim 10$–50~mA for protons and proportionally smaller values for heavy ions, estimated instantaneous luminosities for the $p+\mathrm{Au}$ and $\mathrm{Au}+\mathrm{Au}$ programs are greater than $10^{36}\mathrm{cm^{-2}s^{-1}}$ and $10^{38}\mathrm{cm^{-2}s^{-1}}$, respectively. Figure~\ref{fig:luminosity_fixed_target} shows the available nucleon–nucleon center-of-mass energies $\sqrt{s_{NN}}$ and luminosities for these two configurations.

An example of a seamless coexistence of collider and fixed-target programs is provided by the LHCb collaboration, which studies fixed-target collisions at the LHC with a gas target. Early on, the original System for Measurement of Overlap with Gas (SMOG) operated by injecting gas into the interaction region to function as a target, which also allowed precise determination of beam parameters and luminosity via beam-gas imaging~\cite{LHCb:2014vhh}. The upgraded SMOG2 system consists of a dedicated gas storage cell in front of the LHCb spectrometer, which allows the target mode to operate simultaneously with standard LHC collider running~\cite{BoenteGarcia:2024kba}. This does not require dedicated beam steering and therefore allows large data sets to be collected. The system allows multiple target species to be injected, from H$_2$ to Xe, without detrimental effects on accelerator operations, data acquisition, or reconstruction quality.

\section{Detectors}
\label{sec:section3}

The EIC is designed to accommodate two interaction points (IPs). The ePIC detector will be the first detector, installed at IP6. 
The design of the ePIC detector is compact and largely hermetic, providing nearly full solid-angle coverage and high performance for charged-particle tracking, timing, and particle identification in collider mode. Its geometry and solenoidal magnetic-field configuration, tunable in the range $0.5$–$1.7$~T~\cite{Calvelli:2025vqy} are optimized for the baseline EIC collider program.

In a fixed-target configuration on the hadron-beam side, the target would be inserted along the hadron beam direction, providing $p+A$ and $A+A$ collisions. In this configuration, the center-of-mass frame is strongly boosted relative to the laboratory frame, and most produced particles are emitted in the forward direction with respect to the hadron beam. Consequently, only the forward part of the ePIC detector (hadron-going direction) would provide substantial acceptance for particle reconstruction.

\begin{figure}
\centering
\includegraphics[width=1\linewidth]{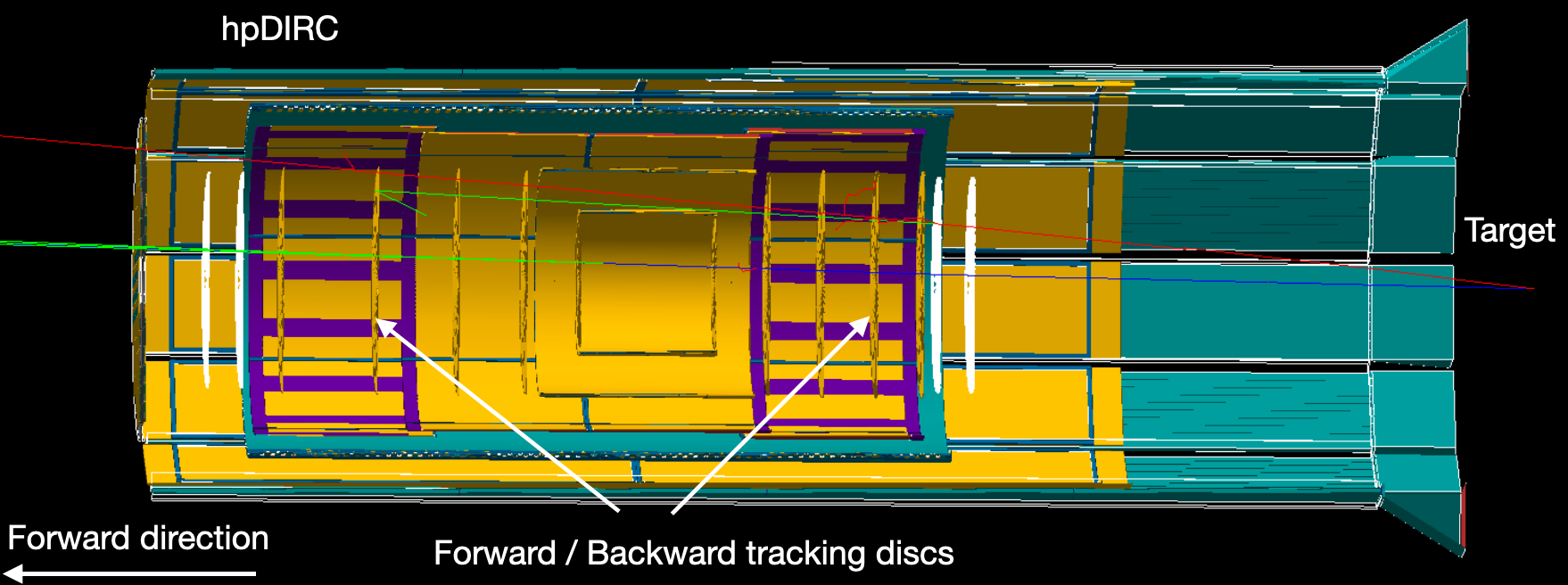}
\caption{
Schematic longitudinal view of a possible ePIC fixed-target configuration based on the ePIC geometry~\cite{ePIC:software}, including the high-performance Detection of Internally Reflected Cherenkov light (hpDIRC) detector, as well as the central and forward tracking detectors. The point-like target is located on the hadron beam side, in the backward region at $z = -3290$~mm with respect to the nominal interaction point (IP6). Fixed-target measurements primarily exploit the forward detector systems.
}
\label{fig:ePIC_FXT_setup}
\end{figure}

The detector layout leaves space available in the backward (lepton-going) region, where a fixed target could be installed with minimal impact on the nominal collider configuration, as illustrated in Fig.~\ref{fig:ePIC_FXT_setup}. This configuration corresponds to a thin foil target, consisting of a very thin layer inserted directly into the beam pipe, as in the STAR FXT program. For the purpose of this study, the target position is set to $z = -3260$~mm; however, this is a provisional placement and may change depending on the final geometry of ePIC and the chosen fixed-target configuration. Such a foil would, in principle, not perturb the circulating beam and could be positioned closer to the central tracking system ($z \sim 0$~mm), substantially enhancing geometrical acceptance, particularly at midrapidity.

The essential requirement is that the target be located downstream of the backward hadronic calorimeter (see Table~\ref{tab:backward_epic_detectors}).

The presence of the Electromagnetic Endcap (EMCal), 2×2×20~cm$^3$ of PbWO$_4$~\cite{Klest:2024acalorimetry}, between the proposed target location and the forward region would not pose an issue. A minimum-ionizing particle (MIP), excluding electrons and photons, loses only about 300~MeV when traversing the EMCal.

The essential detector capabilities for fixed-target operation include:
\begin{itemize}[left=0.1cm]
\item \textbf{Tracking detectors}, providing charged-particle reconstruction and momentum measurement;
\item \textbf{Particle identification (PID) systems}, including fast time-of-flight (TOF) detectors for low-momentum hadron identification ($p \lesssim 2$~GeV) and forward PID detectors, in particular the dual-radiator Ring-Imaging Cherenkov (dRICH) detector, which provides charged-hadron identification up to $p \sim 50$~GeV~\cite{Chatterjee:2024zrn};
\item \textbf{Forward calorimeters}: the forward electromagnetic calorimeter (forward EMCal) could provide muon identification for dimuon measurements, while the forward hadronic calorimeter (forward HCal) would enable hadron measurements~\cite{Klest:2024acalorimetry}.
\end{itemize}

\begin{table}[t]
\centering
\small
\begin{ruledtabular}
\begin{tabular}{lcc} Detector & $z_{\rm start}$ (mm) & $z_{\rm end}$ (mm)\\ \hline HCal (backward) & $-4400$ & $-3725.5$ \\ Backward service gap & $-3460$ & $-3310$ \\ \textbf{Target} & $-3290$ & -- \\ EMCal (backward) & $-2350$ & $-1750$ \\ pfRICH (backward) & $-1685$ & $-1235$ \\ MPGD disks (backward) & $-1225$ & $-1100$ \\ Si disks (backward) & $-1020$ & $-450$ \\ hpDIRC (central) & $-3030$ & $1850$ \\ Si barrel tracker (central) & $-240$ & $240$ \\ Si disks (forward) & $250$ & $1350$ \\ MPGD disks (forward) & $1500$ & $1650$ \\ dRICH (forward) & $1980$ & $3180$ \\ EMCal (forward) & $3350$ & $3620$ \\ HCal (forward) & $3620$ & $4965$ \\ \end{tabular}
\end{ruledtabular}
\caption{
Longitudinal layout of the main ePIC detector components along the beam axis \cite{eic_epic_craterlake}. Negative $z$ values correspond to the lepton-going direction (backward), positive $z$ to the hadron-going direction (forward). The layout includes backward and forward silicon (Si) tracking disks, backward and forward Micro-Pattern Gaseous Detector (MPGD) disks, the central silicon barrel tracker, backward and forward electromagnetic calorimeters (EMCal) and hadronic calorimeters (HCal), the high-performance Detection of Internally Reflected Cherenkov light (hpDIRC) in the barrel, the proximity-focusing Ring-Imaging Cherenkov detector (pfRICH), and the dual-radiator Ring-Imaging Cherenkov detector (dRICH). The solid target at $z=-3290$~mm, shown in Fig.~\ref{fig:ePIC_FXT_setup}, is included as a default reference configuration.}
\label{tab:backward_epic_detectors}
\end{table}

While the fixed-target configuration discussed above demonstrates the flexibility of ePIC (IP6), it represents a baseline approach with inherent limitations in acceptance and optimization for fixed-target physics. To overcome these limitations and fully exploit the scientific opportunities -- such as bridging the energy gap in the QCD phase diagram and providing precise CNM baselines -- a second detector at IP8 could be envisioned, designed from the outset to perform optimally in both collider and fixed-target modes. In this optimized concept, the fixed-target system would be integrated into the baseline layout, featuring a target station with a very-forward detection system immediately downstream. A high-resolution Zero Degree Calorimeter (ZDC) near the target could directly detect spectator neutrons and nuclear remnants, providing robust centrality tagging and access to nuclear breakup and fragmentation observables~\cite{Liu:2022xlm, Kozyrev:2022ehy}. In addition, a precision vertex detector surrounding the interaction region would enable open heavy-flavor measurements by reconstructing displaced decay vertices, significantly enhancing the physics reach of the fixed-target program.
   
\section{Kinematic phase space}
\label{sec:section4}

In this section, we evaluate the preliminary kinematic coverage accessible in a fixed-target configuration at the ePIC detector, considering two observables relevant to the physics program outlined in Sec.~\ref{sec:section1}: $J/\psi$ and charged-pion production. At low collision energies, $\sqrt{s_{NN}} \lesssim 20$~GeV, the formation time of produced partons and pre-hadrons becomes comparable to the nuclear size. This makes the regime particularly sensitive to medium effects acting both at the partonic stage and during hadronization, providing a valuable opportunity to study the interplay between parton propagation and hadron formation in nuclear matter.

\begin{figure}
\centering
\includegraphics[width=1\linewidth]{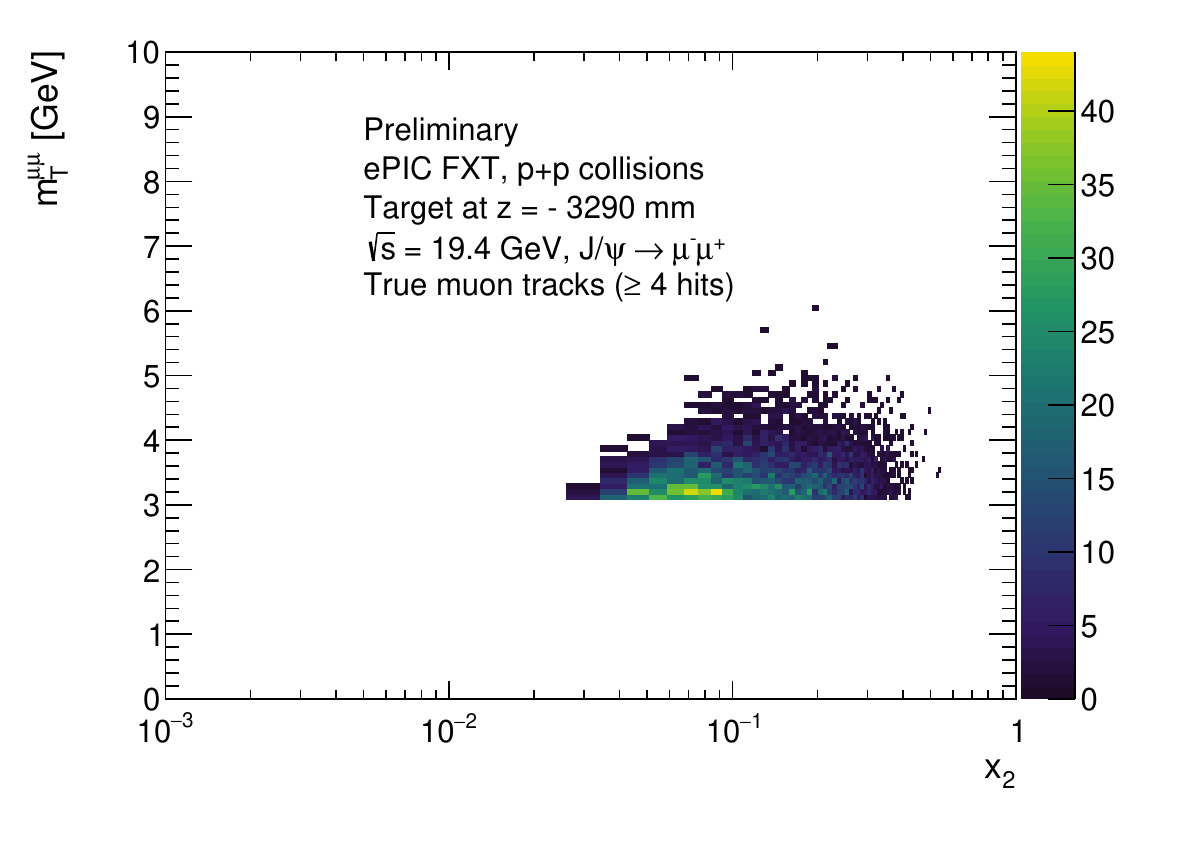}
\caption{
The dimuon transverse mass $m_T^{\mu\mu}$ kinematic phase space as a function of the parton momentum fraction $x_2$ for $J/\psi \rightarrow \mu^{+}\mu^{-}$ decays in fixed-target $p+p$ collisions at ePIC at $\sqrt{s}=19.4$~GeV. The transverse mass is defined as $m_T^{\mu \mu}=\sqrt{M_{\mu \mu}^2+p_T^2}$, where $M_{\mu \mu}$ and $p_T$ denote the invariant mass and transverse momentum of the dimuon, respectively. The simulation uses the nominal magnetic field strength of 1.7~T and the target located at $z = -3290$~mm, as illustrated in Fig.~\ref{fig:ePIC_FXT_setup}. Results are obtained using truth-level muon tracks requiring at least four hits in the tracking detectors.
}
\label{fig:jpsi_kinematics}
\end{figure}

Figure~\ref{fig:jpsi_kinematics} shows the potential kinematic phase space for $J/\psi \rightarrow \mu^{+}\mu^{-}$ decays in $p+p$ collisions at $\sqrt{s} = 19.6$~GeV measured with ePIC, assuming a hypothetical point-like proton target located at $z = -3290$~mm. The $p+p$ simulations serve to illustrate detector acceptance and, for kinematic purposes, are equivalent to $p+A$ ($A+A$) collisions in the absence of nuclear effects. The dimuon transverse mass, $m_T^{\mu\mu}$, is shown as a function of the nuclear (target) momentum fraction $x_2$. The simulation demonstrates sensitivity to transverse momenta $p_T \lesssim M_{J/\psi}$ and to target momentum fractions up to $x_2 \sim 0.4$, covering the nuclear antishadowing and EMC regions~\cite{Eskola:2016oht, Eskola:2021nhw, Kovarik:2015cma, Ball:2022qks}.

\begin{figure}
\centering
\includegraphics[width=1\linewidth]{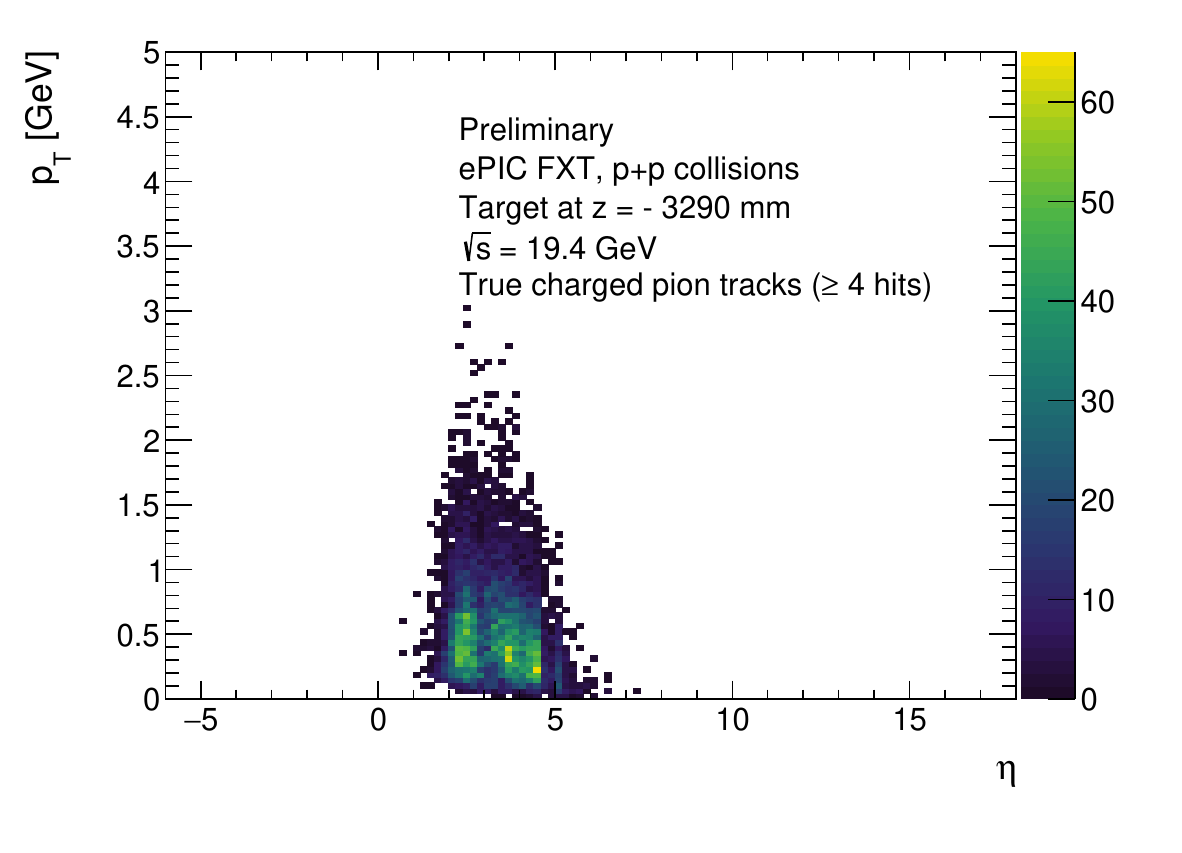}
\caption{
Kinematic phase space of charged-pion transverse momentum $p_T$ as a function of pseudorapidity $\eta$ for charged-pion production in hypothetical fixed-target $p+p$ collisions at ePIC at $\sqrt{s}=19.4$~GeV. The simulation uses the nominal 1.7~T magnetic field and the target positioned at $z = -3290$~mm. Results are obtained using truth-level pion tracks requiring at least four hits in the tracking detectors.
}
\label{fig:pion_kinematics}
\end{figure}

Figure~\ref{fig:pion_kinematics} displays the $p_T$ and pseudorapidity $\eta$ coverage for charged-pion production in the same configuration ($\sqrt{s} = 19.4$~GeV,  point-like proton target at $z = -3290$~mm). Pions are accessible over a broad $\eta$ range (up to $\eta \sim 4$) with $p_T \lesssim 1$~GeV, reflecting the strong forward boost typical of fixed-target kinematics while retaining sensitivity to low-to-moderate-$p_T$ hadrons.

\begin{figure}
\centering
\includegraphics[width=1\linewidth]{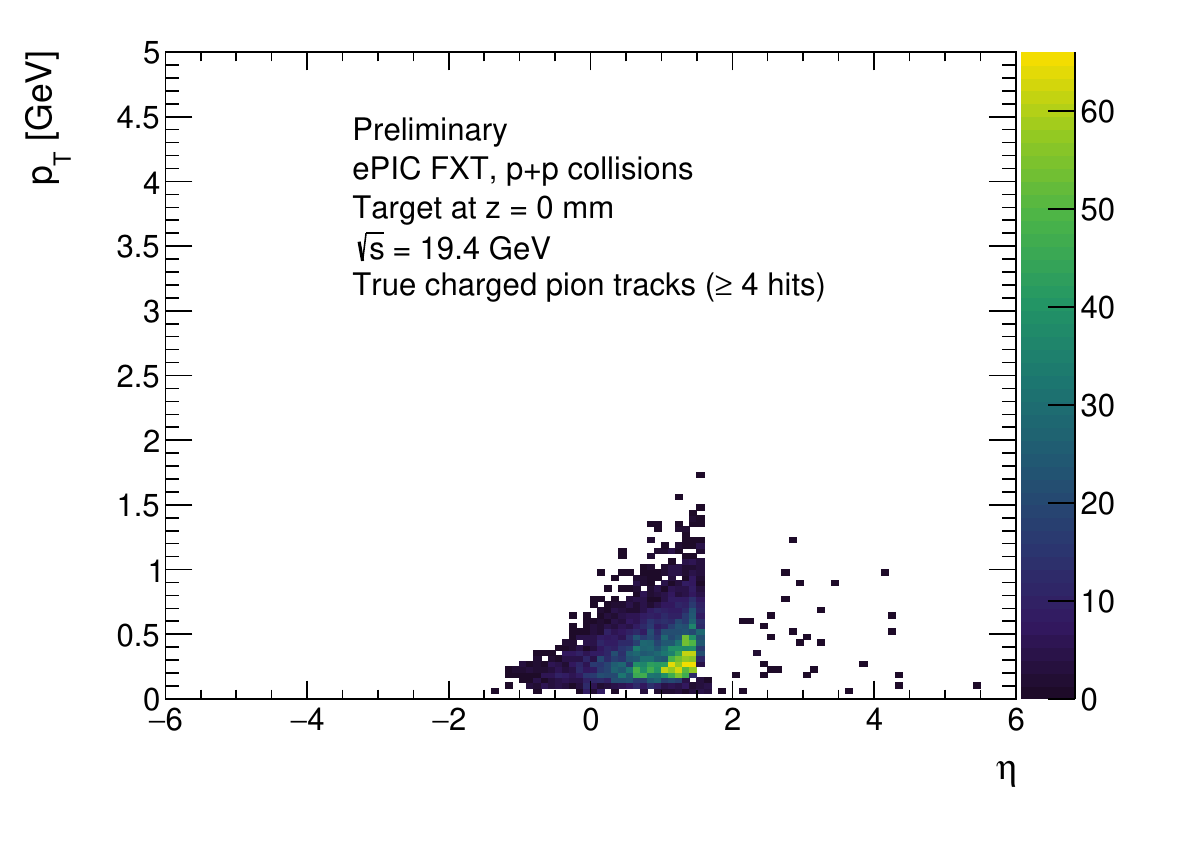}
\caption{
Kinematic phase space of charged-pion transverse momentum $p_T$ as a function of pseudorapidity $\eta$ for charged-pion production in hypothetical fixed-target $p+p$ collisions at ePIC at $\sqrt{s}=19.4$~GeV. The simulation uses the nominal magnetic field strength of 1.7~T and the target positioned at $z = 0$~mm. Results are obtained using truth-level pion tracks requiring at least four hits in the tracking detectors.
}
\label{fig:pion_kinematics_z0}
\end{figure}

Placing a thin foil target closer to the central detector substantially improves midrapidity acceptance. Figure~\ref{fig:pion_kinematics_z0} illustrates this for a target at $z = 0$~mm, enabling charged-pion measurements near $\eta \approx 0$ with comparable $p_T$ coverage and significantly enhanced geometrical acceptance at midrapidity.

The simulations presented here are intended solely to assess the experimental feasibility of fixed-target measurements at the EIC. They demonstrate that the relevant kinematic region at low $\sqrt{s_{NN}}$ can be accessed with the nominal detector configuration, providing a sound starting point for subsequent quantitative studies. 

The choice of the target location directly impacts the physics program. A backward placement is suitable for CNM and space radiation studies, as forward-rapidity particles (large $x_F$) are most relevant. For QCD phase diagram and CP studies, however, midrapidity coverage is essential, as key observables like net-proton cumulants are measured near midrapidity to maximize sensitivity to critical fluctuations. The target position must therefore balance forward and midrapidity requirements to optimize the physics case.    
\section{Conclusion}
\label{sec:conclusion}

The EIC will play a central role in shaping the future of high-energy nuclear physics in the United States. While its primary mission is focused on high-luminosity $e+A$ collisions, the physics potential of the EIC can be significantly broadened through the implementation of a fixed-target program.
Such a fixed-target $p+A$ and $A+A$ program would provide a unique, complementary approach to existing collider-mode measurements, enabling QCD studies in regimes that remain largely unexplored.

By providing high-precision $p+A$ reference data in the $\sqrt{s_{NN}} \sim 10$--20~GeV regime, a fixed-target program at the EIC would establish the missing experimental baselines required to interpret heavy-ion measurements and to disentangle CNM effects from genuine QGP signatures. 
Moreover, the availability of polarized light-ion beams would enable unique studies of spin-dependent nuclear effects.
Such a program would naturally bridge previous fixed-target experiments and the RHIC Beam Energy Scan, enhancing the modern EIC collider program and maximizing its scientific reach. 

In addition, this program would provide valuable nuclear cross-section data in an energy range directly relevant for modeling galactic cosmic-ray interactions with matter, thereby contributing to improved space-radiation transport calculations.

As presented, a baseline fixed-target program could already be realized with the ePIC detector at a relatively low cost, providing immediate scientific value and demonstrating the feasibility of such measurements.
Looking ahead, a future second detector could be designed from the outset to operate optimally in both collider and fixed-target modes. While the program itself would require little additional investment, realizing this long-term potential will depend on early detector design choices that ensure compatibility with fixed-target operation. 

Beyond its direct scientific impact, a fixed-target program provides an opportunity to connect different research communities within the EIC era. By unifying collider and fixed-target capabilities within a single facility, the EIC would emerge not only as a flagship collider but as a comprehensive QCD laboratory, capable of addressing fundamental questions across a broad range of energies, systems, and approaches to studying phases of nuclear matter.
   
\\
\\
{ \large \bf Acknowledgments} \\

We thank the EIC collaboration for providing the ePIC detector simulation framework. The work of C.-J. Naïm is supported by the Center for Frontiers in Nuclear Science and the Simons Foundation. R. Vogt is supported by Lawrence Livermore National Laboratory under Contract DE-AC52-07NA27344 and through the Topical Collaboration in Nuclear Theory on Heavy-Flavor Theory (HEFTY) for QCD Matter under award no. DE-SC0023547.  The work of D. Brown at Brookhaven National Laboratory is sponsored by the Oﬃce of Nuclear Physics, Oﬃce of Science of the U.S. Department of Energy under Contract No. DE-AC02-98CH10886 with Brookhaven Science Associates, LLC.   

\bibliographystyle{apsrev4-2}
\bibliography{biblio.bib}

@misc{Jonas:2026yoz,
    author = "Jonas, Florian and Loizides, Constantin and Mazeliauskas, Aleksas and Paakkinen, Petja and Strangmann, Nicolas",
    title = "{A compendium of cold-nuclear matter baseline predictions in light-ion collisions}",
    eprint = "2602.15928",
    archivePrefix = "arXiv",
    primaryClass = "hep-ph",
    month = "2",
    year = "2026"
}

@article{LHCb:2014rku,
    author = "Aaij, Roel and others",
    collaboration = "LHCb",
    title = "{Study of $\Upsilon$ production and cold nuclear matter effects in $p$Pb collisions at $\sqrt{s_{NN}}$=5 TeV}",
    eprint = "1405.5152",
    archivePrefix = "arXiv",
    primaryClass = "nucl-ex",
    reportNumber = "LHCB-PAPER-2014-015, CERN-PH-EP-2014-102",
    doi = "10.1007/JHEP07(2014)094",
    journal = "JHEP",
    volume = "07",
    pages = "094",
    year = "2014"
}

@article{LHCb:2018psc,
    author = "Aaij, Roel and others",
    collaboration = "LHCb",
    title = "{Study of $\Upsilon$ production in $p$Pb collisions at $\sqrt{s_{NN}}=8.16$ TeV}",
    eprint = "1810.07655",
    archivePrefix = "arXiv",
    primaryClass = "hep-ex",
    reportNumber = "LHCb-PAPER-2018-035, CERN-EP-2018-267",
    doi = "10.1007/JHEP11(2018)194",
    journal = "JHEP",
    volume = "11",
    pages = "194",
    year = "2018",
    note = "[Erratum: JHEP 02, 093 (2020)]"
}

@article{ALICE:2014ict,
    author = "Abelev, Betty Bezverkhny and others",
    collaboration = "ALICE",
    title = "{Production of inclusive $\Upsilon$(1S) and $\Upsilon$(2S) in p-Pb collisions at $\mathbf{\sqrt{s_{{\rm NN}}} = 5.02}$ TeV}",
    eprint = "1410.2234",
    archivePrefix = "arXiv",
    primaryClass = "nucl-ex",
    reportNumber = "CERN-PH-EP-2014-196, ALICE-PUBLIC-2014-002",
    doi = "10.1016/j.physletb.2014.11.041",
    journal = "Phys. Lett. B",
    volume = "740",
    pages = "105--117",
    year = "2015"
}

@article{ALICE:2014cgk,
    author = "Abelev, Betty Bezverkhny and others",
    collaboration = "ALICE",
    title = "{Suppression of $\psi$(2S) production in p-Pb collisions at $\sqrt{s_{\rm NN}}$ = 5.02 TeV}",
    eprint = "1405.3796",
    archivePrefix = "arXiv",
    primaryClass = "nucl-ex",
    reportNumber = "CERN-PH-EP-2014-092",
    doi = "10.1007/JHEP12(2014)073",
    journal = "JHEP",
    volume = "12",
    pages = "073",
    year = "2014"
}

@article{CMS:2018bbk,
    author = "Sirunyan, Albert M. and others",
    collaboration = "CMS",
    title = "{Measurement of exclusive $\Upsilon$ photoproduction from protons in pPb collisions at $\sqrt{s_\mathrm{NN}} =$ 5.02 TeV}",
    eprint = "1809.11080",
    archivePrefix = "arXiv",
    primaryClass = "hep-ex",
    reportNumber = "CMS-FSQ-13-009, CERN-EP-2018-225",
    doi = "10.1140/epjc/s10052-019-6774-8",
    journal = "Eur. Phys. J. C",
    volume = "79",
    number = "3",
    pages = "277",
    year = "2019",
    note = "[Erratum: Eur.Phys.J.C 82, 343 (2022)]"
}

@article{ALICE:2019qie,
    author = "Acharya, Shreyasi and others",
    collaboration = "ALICE",
    title = "{$\Upsilon$ production in pPb collisions at $\sqrt{s_{NN}}$=8.16 TeV}",
    eprint = "1910.14405",
    archivePrefix = "arXiv",
    primaryClass = "nucl-ex",
    reportNumber = "CERN-EP-2019-243",
    doi = "10.1016/j.physletb.2020.135486",
    journal = "Phys. Lett. B",
    volume = "806",
    pages = "135486",
    year = "2020"
}

@article{ALICE:2020vjy,
    author = "Acharya, Shreyasi and others",
    collaboration = "ALICE",
    title = "{Measurement of nuclear effects on $\psi\rm{(2S)}$ production in p-Pb collisions at $\sqrt{\textit{s}_{\rm NN}} = 8.16$ TeV}",
    eprint = "2003.06053",
    archivePrefix = "arXiv",
    primaryClass = "nucl-ex",
    reportNumber = "CERN-EP-2020-036",
    doi = "10.1007/JHEP07(2020)237",
    journal = "JHEP",
    volume = "07",
    pages = "237",
    year = "2020"
}

@article{CMS:2017exb,
    author = "Sirunyan, Albert M and others",
    collaboration = "CMS",
    title = "{Measurement of prompt and nonprompt $\mathrm{J}/{\psi }$ production in $\mathrm {p}\mathrm {p}$ and $\mathrm {p}\mathrm {Pb}$ collisions at $\sqrt{s_{\mathrm {NN}}} =5.02\,\text {TeV} $}",
    eprint = "1702.01462",
    archivePrefix = "arXiv",
    primaryClass = "nucl-ex",
    reportNumber = "CMS-HIN-14-009, CERN-EP-2017-009",
    doi = "10.1140/epjc/s10052-017-4828-3",
    journal = "Eur. Phys. J. C",
    volume = "77",
    number = "4",
    pages = "269",
    year = "2017"
}

@article{LHCb:2024taa,
    author = "Aaij, Roel and others",
    collaboration = "LHCb",
    title = "{Prompt and nonprompt \ensuremath{\psi}(2S) production in pPb collisions at $ \sqrt{s_{\textrm{NN}}} $ = 8.16 TeV}",
    eprint = "2401.11342",
    archivePrefix = "arXiv",
    primaryClass = "hep-ex",
    reportNumber = "LHCb-PAPER-2023-024, CERN-EP-2023-293",
    doi = "10.1007/JHEP04(2024)111",
    journal = "JHEP",
    volume = "04",
    pages = "111",
    year = "2024"
}

@article{LHCb:2016vqr,
    author = "Aaij, Roel and others",
    collaboration = "LHCb",
    title = "{Study of $\psi(2S)$ production and cold nuclear matter effects in pPb collisions at $\sqrt{s_{NN}}=5~\mathrm{TeV}$}",
    eprint = "1601.07878",
    archivePrefix = "arXiv",
    primaryClass = "nucl-ex",
    reportNumber = "LHCB-PAPER-2015-058, CERN-EP-2016-011, LHCb-PAPER-2015-058; CERN-EP-2016-011",
    doi = "10.1007/JHEP03(2016)133",
    journal = "JHEP",
    volume = "03",
    pages = "133",
    year = "2016"
}

@article{LHCb:2013gmv,
    author = "Aaij, R and others",
    collaboration = "LHCb",
    title = "{Study of $J/\psi$ production and cold nuclear matter effects in $pPb$ collisions at $\sqrt{s_{NN}} = 5$ TeV}",
    eprint = "1308.6729",
    archivePrefix = "arXiv",
    primaryClass = "nucl-ex",
    reportNumber = "CERN-PH-EP-2013-156, LHCB-PAPER-2013-052",
    doi = "10.1007/JHEP02(2014)072",
    journal = "JHEP",
    volume = "02",
    pages = "072",
    year = "2014"
}

@article{LHCb:2017ygo,
    author = "Aaij, R. and others",
    collaboration = "LHCb",
    title = "{Prompt and nonprompt J/$\psi$ production and nuclear modification in $p$Pb collisions at $\sqrt{s_{\text{NN}}}= 8.16$ TeV}",
    eprint = "1706.07122",
    archivePrefix = "arXiv",
    primaryClass = "hep-ex",
    reportNumber = "LHCB-PAPER-2017-014, CERN-EP-2017-122",
    doi = "10.1016/j.physletb.2017.09.058",
    journal = "Phys. Lett. B",
    volume = "774",
    pages = "159--178",
    year = "2017"
}

@article{ALICE:2022zig,
    author = "Acharya, Shreyasi and others",
    collaboration = "ALICE",
    title = "{J/$\psi$ production at midrapidity in p$-$Pb collisions at $\sqrt{s_{\rm NN}} = 8.16$ TeV}",
    eprint = "2211.14153",
    archivePrefix = "arXiv",
    primaryClass = "nucl-ex",
    reportNumber = "CERN-EP-2022-256",
    doi = "10.1007/JHEP07(2023)137",
    journal = "JHEP",
    volume = "07",
    pages = "137",
    year = "2023"
}

@article{CMS:2018gbb,
    author = "Sirunyan, Albert M and others",
    collaboration = "CMS",
    title = "{Measurement of prompt $\psi$(2S) production cross sections in proton-lead and proton-proton collisions at $\sqrt{s_{_\mathrm{NN}}}=$ 5.02 TeV}",
    eprint = "1805.02248",
    archivePrefix = "arXiv",
    primaryClass = "hep-ex",
    reportNumber = "CMS-HIN-16-015, CERN-EP-2018-056",
    doi = "10.1016/j.physletb.2019.01.058",
    journal = "Phys. Lett. B",
    volume = "790",
    pages = "509--532",
    year = "2019"
}

@article{BoenteGarcia:2024kba,
    author = "Boente Garcia, O. and others",
    title = "{High-density gas target at the LHCb experiment}",
    eprint = "2407.14200",
    archivePrefix = "arXiv",
    primaryClass = "physics.ins-det",
    reportNumber = "LHCb-DP-2024-002",
    doi = "10.1103/PhysRevAccelBeams.27.111001",
    journal = "Phys. Rev. Accel. Beams",
    volume = "27",
    number = "11",
    pages = "111001",
    year = "2024"
}

@article{Arleo:2020rbm,
    author = "Arleo, François and Naïm, Charles-Joseph",
    title = "{Nuclear $p_\perp$-broadening of Drell-Yan and quarkonium production from SPS to LHC}",
    eprint = "2004.07188",
    archivePrefix = "arXiv",
    primaryClass = "hep-ph",
    doi = "10.1007/JHEP07(2020)220",
    journal = "JHEP",
    volume = "07",
    pages = "220",
    year = "2020"
}

@article{HERA-B:2006bhy,
    author = "Abt, I. and others",
    collaboration = "HERA-B",
    title = "{A Measurement of the $\psi^\prime$ to $J/\psi$ production ratio in 920~GeV proton-nucleus interactions}",
    eprint = "hep-ex/0607046",
    archivePrefix = "arXiv",
    reportNumber = "DESY-06-117",
    doi = "10.1140/epjc/s10052-006-0139-9",
    journal = "Eur. Phys. J. C",
    volume = "49",
    pages = "545--558",
    year = "2007"
}

@article{Arleo:2018zjw,
	Archiveprefix = {arXiv},
	Author = {Arleo, Fran{\c c}ois and Na{\"\i}m, Charles-Joseph and Platchkov, Stephane},
	Doi = {10.1007/JHEP01(2019)129},
	Eprint = {1810.05120},
	Journal = {JHEP},
	Pages = {129},
	Primaryclass = {hep-ph},
	Slaccitation = {%%CITATION = ARXIV:1810.05120;%%},
	Title = {{Initial-state energy loss in cold QCD matter and the Drell-Yan process}},
	Volume = {01},
	Year = {2019}}

@phdthesis{Lin:2017eoc,
	Author = {Lin, Po-Ju},
	Doi = {10.2172/1398791},
	Reportnumber = {FERMILAB-THESIS-2017-18},
	School = {Colorado U.},
	Slaccitation = {%%CITATION = FERMILAB-THESIS-2017-18;%%},
	Title = {{Measurement of Quark Energy Loss in Cold Nuclear Matter at Fermilab E906/SeaQuest}},
	Url = {\url{http://lss.fnal.gov/archive/thesis/2000/fermilab-thesis-2017-18.pdf}},
	Year = {2017}}

@article{Kovarik:2015cma,
	Archiveprefix = {arXiv},
	Author = {Kovarik, K. and others},
	Doi = {10.1103/PhysRevD.93.085037},
	Eprint = {1509.00792},
	Journal = {Phys. Rev.},
	Pages = {085037},
	Primaryclass = {hep-ph},
	Reportnumber = {LPSC-15-153, MS-TP-15-11, FERMILAB-PUB-15-375-ND-PPD-T},
	Slaccitation = {%%CITATION = ARXIV:1509.00792;%%},
	Title = {{nCTEQ15 - Global analysis of nuclear parton distributions with uncertainties in the CTEQ framework}},
	Volume = {D93},
	Year = {2016}}

@article{Adare:2012qf,
	Archiveprefix = {arXiv},
	Author = {Adare, A. and others},
	Collaboration = {PHENIX},
	Doi = {10.1103/PhysRevC.87.034904},
	Eprint = {1204.0777},
	Journal = {Phys. Rev.},
	Number = {3},
	Pages = {034904},
	Primaryclass = {nucl-ex},
	Slaccitation = {%%CITATION = ARXIV:1204.0777;%%},
	Title = {{Transverse-Momentum Dependence of the $J/\psi$ Nuclear Modification in dAu Collisions at $\sqrt{s_{NN}}=200$ GeV}},
	Volume = {C87},
	Year = {2013}}

@article{Arleo:2012hn,
	Archiveprefix = {arXiv},
	Author = {Arleo, Fran{\c c}ois and Peign{\'e}, St{\'e}phane},
	Doi = {10.1103/PhysRevLett.109.122301},
	Eprint = {1204.4609},
	Journal = {Phys. Rev. Lett.},
	Pages = {122301},
	Primaryclass = {hep-ph},
	Reportnumber = {CERN-PH-TH-2012-116},
	Slaccitation = {%%CITATION = ARXIV:1204.4609;%%},
	Title = {{J/$\psi$ suppression in pA collisions from parton energy loss in cold QCD matter}},
	Volume = {109},
	Year = {2012}}

@article{Adam:2015jsa,
      author         = "Adam, Jaroslav and others",
      title          = "{Centrality dependence of inclusive J/$\psi$ production in
                        p-Pb collisions at $ \sqrt{s_{\mathrm{NN}}}=5.02 $ TeV}",
      collaboration  = "ALICE",
      journal        = "JHEP",
      volume         = "11",
      year           = "2015",
      pages          = "127",
      doi            = "10.1007/JHEP11(2015)127",
      eprint         = "1506.08808",
      archivePrefix  = "arXiv",
      primaryClass   = "nucl-ex",
      reportNumber   = "CERN-PH-EP-2015-158",
      SLACcitation   = "%%CITATION = ARXIV:1506.08808;%%"
}

@article{STAR:2021zvb,
    author = "Abdallah, Mohamed and others",
    collaboration = "STAR",
    title = "{Measurement of cold nuclear matter effects for inclusive J/\ensuremath{\psi} in p+Au collisions at sNN=200 GeV}",
    eprint = "2110.09666",
    archivePrefix = "arXiv",
    primaryClass = "nucl-ex",
    doi = "10.1016/j.physletb.2021.136865",
    journal = "Phys. Lett. B",
    volume = "825",
    pages = "136865",
    year = "2022"
}

@article{LHCb:2022sxs,
    author = "Aaij, R. and others",
    collaboration = "LHCb",
    title = "{Charmonium production in $p$Ne collisions at $\sqrt{s_{\scriptscriptstyle \text {NN}}} =68.5$ GeV}",
    eprint = "2211.11645",
    archivePrefix = "arXiv",
    primaryClass = "hep-ex",
    reportNumber = "CERN-EP-2022-216, LHCb-PAPER-2022-014",
    doi = "10.1140/epjc/s10052-023-11608-6",
    journal = "Eur. Phys. J. C",
    volume = "83",
    number = "7",
    pages = "625",
    year = "2023"
}

@article{STAR:2013kwk,
    author = "Adamczyk, L. and others",
    collaboration = "STAR",
    title = "{Suppression of $\Upsilon$ production in d+Au and Au+Au collisions at $\sqrt{s_{NN}}$=200 GeV}",
    eprint = "1312.3675",
    archivePrefix = "arXiv",
    primaryClass = "nucl-ex",
    doi = "10.1016/j.physletb.2014.06.028",
    journal = "Phys. Lett. B",
    volume = "735",
    pages = "127--137",
    year = "2014",
    note = "[Erratum: Phys.Lett.B 743, 537--541 (2015)]"
}

@article{STAR:2023nos,
    author = "Abdulhamid, M. I. and others",
    collaboration = "STAR",
    title = "{Observation of Strong Nuclear Suppression in Exclusive J/\ensuremath{\psi} Photoproduction in Au+Au Ultraperipheral Collisions at RHIC}",
    eprint = "2311.13637",
    archivePrefix = "arXiv",
    primaryClass = "nucl-ex",
    doi = "10.1103/PhysRevLett.133.052301",
    journal = "Phys. Rev. Lett.",
    volume = "133",
    number = "5",
    pages = "052301",
    year = "2024"
}

@article{Odyniec:2013aaa,
    author = "Odyniec, Grazyna",
    title = "{RHIC Beam Energy Scan Program: Phase I and II}",
    doi = "10.22323/1.185.0043",
    journal = "PoS",
    volume = "CPOD2013",
    pages = "043",
    year = "2013"
}

@article{Ozawa:2022sam,
    author = "Ozawa, Kyoichiro and others",
    title = "{The J-PARC heavy ion project}",
    doi = "10.1051/epjconf/202227111004",
    journal = "EPJ Web Conf.",
    volume = "271",
    pages = "11004",
    year = "2022"
}

@article{STAR:2013gus,
    author = "Adamczyk, L. and others",
    collaboration = "STAR",
    title = "{Energy Dependence of Moments of Net-proton Multiplicity Distributions at RHIC}",
    eprint = "1309.5681",
    archivePrefix = "arXiv",
    primaryClass = "nucl-ex",
    doi = "10.1103/PhysRevLett.112.032302",
    journal = "Phys. Rev. Lett.",
    volume = "112",
    pages = "032302",
    year = "2014"
}

@mis{LHCb:2025ixz,
    author = "Aaij, Roel and others",
    collaboration = "LHCb",
    title = "{Unveiling the shape of the $^{20}$Ne nucleus by measuring the flow coefficients with cumulants in PbNe and PbAr collisions at $\sqrt{s_{NN}} = 70.9$ GeV}",
    eprint = "2509.12399",
    archivePrefix = "arXiv",
    primaryClass = "nucl-ex",
    reportNumber = "LHCb-CONF-2025-001",
    month = "9",
    year = "2025"
}

@article{LHCb:2022qvj,
    author = "Aaij, R. and others",
    collaboration = "LHCb",
    title = "{${{J}/\psi }$ and ${{D}} ^0$ production in $\sqrt{s_{\scriptscriptstyle \text {NN}}} =68.5\,\text {GeV} $ PbNe collisions}",
    eprint = "2211.11652",
    archivePrefix = "arXiv",
    primaryClass = "hep-ex",
    reportNumber = "CERN-EP-2022-215, LHCb-PAPER-2022-011",
    doi = "10.1140/epjc/s10052-023-11674-w",
    journal = "Eur. Phys. J. C",
    volume = "83",
    number = "7",
    pages = "658",
    year = "2023"
}

@article{Baranov:2018cdk,
    author = "Baranov, D. and Kapishin, M. and Mamontova, T. and Pokatashkin, G. and Rufanov, I. and Vasendina, V. and Zinchenko, A.",
    title = "{The BM@N Experiment at JINR: Status and Physics Program}",
    doi = "10.18502/ken.v3i1.1757",
    journal = "KnE Energ. Phys.",
    volume = "3",
    pages = "291--296",
    year = "2018"
}

@article{Vogt:2011zzb,
    author = "Vogt, R. and Lourenco, C. and Wohri, H. K.",
    editor = "Tserruya, Itzhak and Milov, Alexander and d'Enterria, David and Jacobs, Peter and Wiedemann, Urs",
    title = "{J/psi production and absorption in p + A and d + Au collisions}",
    doi = "10.1016/j.nuclphysa.2011.02.104",
    journal = "Nucl. Phys. A",
    volume = "855",
    pages = "453--456",
    year = "2011"
}

@article{Liu:2022xlm,
    author = "Liu, Lu-Meng and Zhang, Chun-Jian and Xu, Jun and Jia, Jiangyong and Peng, Guang-Xiong",
    title = "{Free spectator nucleons in ultracentral relativistic heavy-ion collisions as a probe of neutron skin}",
    eprint = "2209.03106",
    archivePrefix = "arXiv",
    primaryClass = "nucl-th",
    doi = "10.1103/PhysRevC.106.034913",
    journal = "Phys. Rev. C",
    volume = "106",
    number = "3",
    pages = "034913",
    year = "2022"
}

@article{Kozyrev:2022ehy,
    author = "Kozyrev, Nikita and Svetlichnyi, Aleksandr and Nepeivoda, Roman and Pshenichnov, Igor",
    title = "{Peeling away neutron skin in ultracentral collisions of relativistic nuclei}",
    eprint = "2204.07189",
    archivePrefix = "arXiv",
    primaryClass = "nucl-th",
    doi = "10.1140/epja/s10050-022-00832-5",
    journal = "Eur. Phys. J. A",
    volume = "58",
    number = "9",
    pages = "184",
    year = "2022"
}

@article{Vogt:2021vsc,
    author = "Vogt, R.",
    title = "{Limits on Intrinsic Charm Production from the SeaQuest Experiment}",
    eprint = "2101.02858",
    archivePrefix = "arXiv",
    primaryClass = "hep-ph",
    reportNumber = "LLNL-JRNL-818026",
    doi = "10.1103/PhysRevC.103.035204",
    journal = "Phys. Rev. C",
    volume = "103",
    number = "3",
    pages = "035204",
    year = "2021"
}

@article{Vogt:2023plx,
    author = "Vogt, R.",
    title = "{Contribution from intrinsic charm production to fixed-target interactions with the SMOG Device at LHCb}",
    eprint = "2304.09356",
    archivePrefix = "arXiv",
    primaryClass = "hep-ph",
    reportNumber = "LLNL-JRNL-847208",
    doi = "10.1103/PhysRevC.108.015201",
    journal = "Phys. Rev. C",
    volume = "108",
    number = "1",
    pages = "015201",
    year = "2023"
}

@inproceedings{Liu:2021ypa,
    author = "Liu, Chuyu and others",
    title = "{Review of the Fixed Target Operation at RHIC in 2020}",
    booktitle = "{12th International Particle Accelerator Conference~}",
    doi = "10.18429/JACoW-IPAC2021-MOPAB009",
    month = "8",
    year = "2021"
}

@article{Smith:2023fspas,
author       = {Smith, Michael S. and Vogt, Ramona L. and LaBel, Kenneth A.},
title        = {Nuclear data for space exploration},
journal      = {Frontiers in Astronomy and Space Sciences},
volume       = {10},
year         = {2023},
pages        = {1228901},
doi          = {10.3389/fspas.2023.1228901},
url          = {https://www.frontiersin.org/articles/10.3389/fspas.2023.1228901}
}

@article{10.1371/journal.pbio.3000669,
    doi = {10.1371/journal.pbio.3000669},
    author = {Simonsen, Lisa C. AND Slaba, Tony C. AND Guida, Peter AND Rusek, Adam},
    journal = {PLOS Biology},
    publisher = {Public Library of Science},
    title = {NASA’s first ground-based Galactic Cosmic Ray Simulator: Enabling a new era in space radiobiology research},
    year = {2020},
    month = {05},
    volume = {18},
    url = {https://doi.org/10.1371/journal.pbio.3000669},
    pages = {1-32},
    number = {5},

}

@article{Chancellor03042021,
author = {Jeffery Chancellor and Craig Nowadly and Jacqueline Williams and Serena Aunon-Chancellor and Megan Chesal and Jayme Looper and Wayne Newhauser},
title = {Everything you wanted to know about space radiation but were afraid to ask},
journal = {Journal of Environmental Science and Health, Part C},
volume = {39},
number = {2},
pages = {113--128},
year = {2021},
publisher = {Taylor \& Francis},
doi = {10.1080/26896583.2021.1897273},
note ={PMID: 33902392},
URL = {https://doi.org/10.1080/26896583.2021.1897273},
eprint = {https://doi.org/10.1080/26896583.2021.1897273}
}

@misc{luoni2021totalnuclearreactioncrosssection,
      title={Total nuclear reaction cross-section database for radiation protection in space and heavy-ion therapy applications}, 
      author={F Luoni and F Horst and CA Reidel and A Quarz and L Bagnale and L Sihver and U Weber and RB Norman and W de Wet and M Giraudo and G Santin and J Norbury and M Durante},
      year={2021},
      eprint={2105.11981},
      archivePrefix={arXiv},
      primaryClass={nucl-th},
      doi={https://doi.org/10.1088/1367-2630/ac27e1},
      url={https://arxiv.org/abs/2105.11981}, 
}

@article{NORBURY2012315,
title = {Nuclear data for space radiation},
journal = {Radiation Measurements},
volume = {47},
number = {5},
pages = {315-363},
year = {2012},
issn = {1350-4487},
doi = {https://doi.org/10.1016/j.radmeas.2012.03.004},
url = {https://www.sciencedirect.com/science/article/pii/S1350448712000686},
author = {John W. Norbury and Jack Miller and Anne M. Adamczyk and Lawrence H. Heilbronn and Lawrence W. Townsend and Steve R. Blattnig and Ryan B. Norman and Stephen B. Guetersloh and Cary J. Zeitlin}
}

@article{STAR:2019bjj,
    author = "Adam, Jaroslav and others",
    collaboration = "STAR",
    title = "{Strange hadron production in Au+Au collisions at $\sqrt{s_{_{\mathrm{NN}}}}$ = 7.7, 11.5, 19.6, 27, and 39 GeV}",
    eprint = "1906.03732",
    archivePrefix = "arXiv",
    primaryClass = "nucl-ex",
    doi = "10.1103/PhysRevC.102.034909",
    journal = "Phys. Rev. C",
    volume = "102",
    number = "3",
    pages = "034909",
    year = "2020"
}

@article{Agarwal:2023otg,
    author = "Agarwal, Kshitij",
    collaboration = "CBM",
    title = "{The compressed baryonic matter (CBM) experiment at FAIR{\textemdash}physics, status and prospects}",
    doi = "10.1088/1402-4896/acbca7",
    journal = "Phys. Scripta",
    volume = "98",
    number = "3",
    pages = "034006",
    year = "2023"
}

@misc{eic_epic_craterlake,
  title        = {{epic\_craterlake.csv}},
  author       = {{EIC Collaboration}},
  howpublished = {\url{https://eic.github.io/epic/artifacts/DetectorParameterTable/epic_craterlake.csv}},
  note         = {Detector parameter table for the ePIC geometry description},
  year         = {2026},
  key          = {epic_craterlake}
}

@article{Klest:2024acalorimetry,
    author = "Klest, Henry T.",
    collaboration = "ePIC",
    title = "{Calorimetry for the ePIC Experiment}",
    eprint = "2408.11075",
    archivePrefix = "arXiv",
    primaryClass = "physics.ins-det",
    doi = "10.22323/1.469.0276",
    journal = "PoS",
    volume = "DIS2024",
    pages = "276",
    year = "2025"
}

@misc{ePIC:software,
  title        = {{ePIC} detector simulation software},
  author       = {{ePIC Collaboration}},
  howpublished = {\url{https://github.com/eic/epic}},
  year         = {2024},
  note         = {Public GitHub repository, accessed January 2026}
}

@article{Neufeld:2010dz,
	Archiveprefix = {arXiv},
	Author = {Neufeld, R.B. and Vitev, Ivan and Zhang, Ben-Wei},
	Doi = {10.1016/j.physletb.2011.09.045},
	Eprint = {1010.3708},
	Journal = {Phys.Lett.},
	Pages = {590},
	Primaryclass = {hep-ph},
	Slaccitation = {%%CITATION = ARXIV:1010.3708;%%},
	Title = {{A possible determination of the quark radiation length in cold nuclear matter}},
	Volume = {B704},
	Year = {2011}}

@article{deFlorian:2011fp,
	Archiveprefix = {arXiv},
	Author = {de Florian, Daniel and Sassot, Rodolfo and Zurita, Pia and Stratmann, Marco},
	Doi = {10.1103/PhysRevD.85.074028},
	Eprint = {1112.6324},
	Journal = {Phys.\ Rev.},
	Pages = {074028},
	Primaryclass = {hep-ph},
	Slaccitation = {%%CITATION = ARXIV:1112.6324;%%},
	Title = {{Global Analysis of Nuclear Parton Distributions}},
	Volume = {D85},
	Year = {2012}}

@article{Alde:1991sw,
      author         = "Alde, D. M. and others",
      title          = "{Nuclear dependence of the production of Upsilon
                        resonances at 800-GeV}",
      journal        = "Phys. Rev. Lett.",
      volume         = "66",
      year           = "1991",
      pages          = "2285-2288",
      doi            = "10.1103/PhysRevLett.66.2285",
      reportNumber   = "FERMILAB-PUB-91-037-E",
      SLACcitation   = "%%CITATION = PRLTA,66,2285;%%"
}

@article{Branson:1977ci,
    author = "Branson, J. G. and Sanders, G. H. and Smith, A. J. S. and Thaler, J. J and Anderson, K. J. and Henry, G. G and McDonald, K. T. and Pilcher, J. E. and Rosenberg, E. I.",
    title = "{Production of the $J/\psi$ and $\psi^\prime$ (3.7) by 225~GeV/c $\pi^{\pm}$ and Proton Beams on C and Sn Targets.}",
    reportNumber = "FERMILAB-PUB-77-177-E, D77-06182",
    doi = "10.1103/PhysRevLett.38.1331",
    journal = "Phys. Rev. Lett.",
    volume = "38",
    pages = "1331--1334",
    year = "1977"
}

@article{Antipov:1977ss,
    author = "Antipov, Yu. M. and others",
    title = "{J/psi Production off Be, Cu and W by 43-GeV/c Negative Pions}",
    doi = "10.1016/0370-2693(77)90721-3",
    journal = "Phys. Lett. B",
    volume = "72",
    pages = "278--280",
    year = "1977"
}

@article{NA50:2006rdp,
    author = "Alessandro, B. and others",
    collaboration = "NA50",
    title = "{J/psi and psi-prime production and their normal nuclear absorption in proton-nucleus collisions at 400-GeV}",
    eprint = "nucl-ex/0612012",
    archivePrefix = "arXiv",
    reportNumber = "CERN-PH-EP-2006-018",
    doi = "10.1140/epjc/s10052-006-0079-4",
    journal = "Eur. Phys. J. C",
    volume = "48",
    pages = "329",
    year = "2006"
}

@article{Badier:1983dg,
    author = "Badier, J. and others",
    collaboration = "NA3",
    title = "{Experimental J/psi Hadronic Production from 150-GeV/c to 280-GeV/c}",
    reportNumber = "CERN-EP/83-86",
    doi = "10.1007/BF01573213",
    journal = "Z. Phys. C",
    volume = "20",
    pages = "101",
    year = "1983"
}

@article{Fu:2019hdw,
    author = "Fu, Wei-jie and Pawlowski, Jan M. and Rennecke, Fabian",
    title = "{QCD phase structure at finite temperature and density}",
    eprint = "1909.02991",
    archivePrefix = "arXiv",
    primaryClass = "hep-ph",
    doi = "10.1103/PhysRevD.101.054032",
    journal = "Phys. Rev. D",
    volume = "101",
    number = "5",
    pages = "054032",
    year = "2020"
}

@article{Schmidt:2022ogw,
    author = "Schmidt, Christian and Clarke, David A. and Nicotra, Guido and Di Renzo, Francesco and Dimopoulos, Petros and Singh, Simran and Goswami, Jishnu and Zambello, Kevin",
    title = "{Detecting Critical Points from the Lee{\textendash}Yang Edge Singularities in Lattice QCD}",
    eprint = "2209.04345",
    archivePrefix = "arXiv",
    primaryClass = "hep-lat",
    doi = "10.5506/APhysPolBSupp.16.1-A52",
    journal = "Acta Phys. Polon. Supp.",
    volume = "16",
    number = "1",
    pages = "1--A52",
    year = "2023"
}

@article{Zambello:2023ptp,
    author = "Zambello, Kevin and Clarke, David Anthony and Dimopoulos, Petros and Di Renzo, Francesco and Goswami, Jishnu and Nicotra, Guido and Schmidt, Christian and Singh, Simran",
    title = "{Determination of Lee-Yang edge singularities in QCD by rational approximations}",
    eprint = "2301.03952",
    archivePrefix = "arXiv",
    primaryClass = "hep-lat",
    doi = "10.22323/1.430.0164",
    journal = "PoS",
    volume = "LATTICE2022",
    pages = "164",
    year = "2023"
}

@article{Basar:2023nkp,
    author = "Basar, Gokce",
    title = "{QCD critical point, Lee-Yang edge singularities, and Pad{\'e} resummations}",
    eprint = "2312.06952",
    archivePrefix = "arXiv",
    primaryClass = "hep-th",
    doi = "10.1103/PhysRevC.110.015203",
    journal = "Phys. Rev. C",
    volume = "110",
    number = "1",
    pages = "015203",
    year = "2024"
}

@article{Clarke:2024ugt,
    author = "Clarke, David A. and Dimopoulos, Petros and Di Renzo, Francesco and Goswami, Jishnu and Schmidt, Christian and Singh, Simran and Zambello, Kevin",
    title = "{Searching for the QCD critical end point using multipoint Pad{\'e} approximations}",
    eprint = "2405.10196",
    archivePrefix = "arXiv",
    primaryClass = "hep-lat",
    doi = "10.1103/y6kg-ry8x",
    journal = "Phys. Rev. D",
    volume = "112",
    number = "9",
    pages = "L091504",
    year = "2025"
}

@article{Hippert:2023bel,
    author = "Hippert, Mauricio and Grefa, Joaquin and Manning, T. Andrew and Noronha, Jorge and Noronha-Hostler, Jacquelyn and Portillo Vazquez, Israel and Ratti, Claudia and Rougemont, Romulo and Trujillo, Michael",
    title = "{Bayesian location of the QCD critical point from a holographic perspective}",
    eprint = "2309.00579",
    archivePrefix = "arXiv",
    primaryClass = "nucl-th",
    doi = "10.1103/PhysRevD.110.094006",
    journal = "Phys. Rev. D",
    volume = "110",
    number = "9",
    pages = "094006",
    year = "2024"
}

@article{Shah:2024img,
    author = "Shah, Hitansh and Hippert, Mauricio and Noronha, Jorge and Ratti, Claudia and Vovchenko, Volodymyr",
    title = "{Locating the QCD critical point through contours of constant entropy density}",
    eprint = "2410.16206",
    archivePrefix = "arXiv",
    primaryClass = "hep-ph",
    doi = "10.1103/cbwj-4jfl",
    journal = "Phys. Rev. C",
    volume = "113",
    number = "1",
    pages = "L012201",
    year = "2026"
}

@misc{STAR:2025rpo,
    collaboration = "STAR",
    title = "{Charged-Particle Multiplicity Dependence of Net-Proton Cumulants in Zr+Zr and Ru+Ru Collisions at $\sqrt{s_{NN}}$ = 200 GeV}",
    eprint = "2511.05043",
    archivePrefix = "arXiv",
    primaryClass = "nucl-ex",
    month = "11",
    year = "2025"
}

@misc{ALICE:2026ajq,
    author = "Abualrob, Ibrahim Jaser and others",
    collaboration = "ALICE",
    title = "{Inclusive and multiplicity-dependent pseudorapidity densities of charged particles in pp collisions at $\mathbf{\sqrt{s} = 13.6}$ TeV}",
    eprint = "2602.10658",
    archivePrefix = "arXiv",
    primaryClass = "nucl-ex",
    reportNumber = "CERN-EP-2026-012",
    month = "2",
    year = "2026"
}

@article{STAR:2023zhl,
    author = "Abdulhamid, Muhammad and others",
    collaboration = "STAR",
    title = "{Measurements of charged-particle multiplicity dependence of higher-order net-proton cumulants in p+p collisions at s=200 GeV from STAR at RHIC}",
    eprint = "2311.00934",
    archivePrefix = "arXiv",
    primaryClass = "nucl-ex",
    doi = "10.1016/j.physletb.2024.138966",
    journal = "Phys. Lett. B",
    volume = "857",
    pages = "138966",
    year = "2024"
}

@article{STAR:2021fge,
    author = "Abdallah, M. S. and others",
    collaboration = "STAR",
    title = "{Measurements of Proton High Order Cumulants in $\sqrt{s_{_{\mathrm{NN}}}}$ = 3 GeV Au+Au Collisions and Implications for the QCD Critical Point}",
    eprint = "2112.00240",
    archivePrefix = "arXiv",
    primaryClass = "nucl-ex",
    doi = "10.1103/PhysRevLett.128.202303",
    journal = "Phys. Rev. Lett.",
    volume = "128",
    number = "20",
    pages = "202303",
    year = "2022"
}

@article{STAR:2022etb,
    author = "Abdallah, Mohamed and others",
    collaboration = "STAR",
    title = "{Higher-order cumulants and correlation functions of proton multiplicity distributions in sNN=3~GeV~Au+Au collisions at the RHIC STAR experiment}",
    eprint = "2209.11940",
    archivePrefix = "arXiv",
    primaryClass = "nucl-ex",
    doi = "10.1103/PhysRevC.107.024908",
    journal = "Phys. Rev. C",
    volume = "107",
    number = "2",
    pages = "024908",
    year = "2023"
}

@article{STAR:2024znc,
    author = "Abdulhamid, M. I. and others",
    collaboration = "STAR",
    title = "{Strangeness production in $ \sqrt{s_{\textrm{NN}}} $ = 3 GeV Au+Au collisions at RHIC}",
    eprint = "2407.10110",
    archivePrefix = "arXiv",
    primaryClass = "nucl-ex",
    doi = "10.1007/JHEP10(2024)139",
    journal = "JHEP",
    volume = "10",
    pages = "139",
    year = "2024"
}

@article{STAR:2024zvj,
    author = "Aboona, B. E. and others",
    collaboration = "STAR",
    title = "{Light nuclei femtoscopy and baryon interactions in 3 GeV Au+Au collisions at RHIC}",
    eprint = "2410.03436",
    archivePrefix = "arXiv",
    primaryClass = "nucl-ex",
    doi = "10.1016/j.physletb.2025.139412",
    journal = "Phys. Lett. B",
    volume = "864",
    pages = "139412",
    year = "2025"
}

@misc{Sorensen:2024mry,
    author = "Sorensen, Agnieszka and Sorensen, Paul",
    title = "{Locating the critical point for the hadron to quark-gluon plasma phase transition from finite-size scaling of proton cumulants in heavy-ion collisions}",
    eprint = "2405.10278",
    archivePrefix = "arXiv",
    primaryClass = "nucl-th",
    reportNumber = "INT-PUB-24-018",
    month = "5",
    year = "2024"
}

@article{Gao:2020fbl,
    author = "Gao, Fei and Pawlowski, Jan M.",
    title = "{Chiral phase structure and critical end point in QCD}",
    eprint = "2010.13705",
    archivePrefix = "arXiv",
    primaryClass = "hep-ph",
    doi = "10.1016/j.physletb.2021.136584",
    journal = "Phys. Lett. B",
    volume = "820",
    pages = "136584",
    year = "2021"
}

@article{Mukherjee:2019eou,
    author = "Mukherjee, Swagato and Skokov, Vladimir",
    title = "{Universality driven analytic structure of the QCD crossover: radius of convergence in the baryon chemical potential}",
    eprint = "1909.04639",
    archivePrefix = "arXiv",
    primaryClass = "hep-ph",
    doi = "10.1103/PhysRevD.103.L071501",
    journal = "Phys. Rev. D",
    volume = "103",
    number = "7",
    pages = "L071501",
    year = "2021"
}

@article{STAR:2016vqt,
    author = "Adamczyk, L. and others",
    collaboration = "STAR",
    title = "{Beam Energy Dependence of the Third Harmonic of Azimuthal Correlations in Au+Au Collisions at RHIC}",
    eprint = "1601.01999",
    archivePrefix = "arXiv",
    primaryClass = "nucl-ex",
    doi = "10.1103/PhysRevLett.116.112302",
    journal = "Phys. Rev. Lett.",
    volume = "116",
    number = "11",
    pages = "112302",
    year = "2016"
}

@article{Du:2024wjm,
    author = "Du, Lipei and Sorensen, Agnieszka and Stephanov, Mikhail",
    title = "{The QCD phase diagram and Beam Energy Scan physics: A theory overview}",
    eprint = "2402.10183",
    archivePrefix = "arXiv",
    primaryClass = "nucl-th",
    reportNumber = "INT-PUB-24-017",
    doi = "10.1142/9789811294679_0007",
    journal = "Int. J. Mod. Phys. E",
    volume = "33",
    number = "07",
    pages = "2430008",
    year = "2024"
}

@article{Sorensen:2023zkk,
    author = "Sorensen, Agnieszka and others",
    title = "{Dense nuclear matter equation of state from heavy-ion collisions}",
    eprint = "2301.13253",
    archivePrefix = "arXiv",
    primaryClass = "nucl-th",
    reportNumber = "INT-PUB-23-001, LA-UR-23-20514, LLNL-TR-844629",
    doi = "10.1016/j.ppnp.2023.104080",
    journal = "Prog. Part. Nucl. Phys.",
    volume = "134",
    pages = "104080",
    year = "2024"
}

@misc{RHICRunOverview,
  title        = {RHIC Run Overview},
  author       = {{Brookhaven National Laboratory}},
  howpublished = {\url{https://www.agsrhichome.bnl.gov/RHIC/Runs/}},
  note         = {Accessed: 2026-02-21},
  year         = {2026}
}

@article{Leitch:1999ea,
    author = "Leitch, M. J. and others",
    collaboration = "NuSea",
    title = "{Measurement of J / psi and psi-prime suppression in p-A collisions at 800-GeV/c}",
    eprint = "nucl-ex/9909007",
    archivePrefix = "arXiv",
    reportNumber = "LA-UR-99-5007, FERMILAB-PUB-99-394-E",
    doi = "10.1103/PhysRevLett.84.3256",
    journal = "Phys. Rev. Lett.",
    volume = "84",
    pages = "3256--3260",
    year = "2000"
}

@article{Calvelli:2025vqy,
    author = "Calvelli, Valerio and others",
    title = "{Design of MARCO, the New Solenoidal Detector Magnet for the ePIC Experiment at BNL}",
    reportNumber = "JLAB-PHY-25-4272",
    doi = "10.1109/TASC.2025.3608471",
    journal = "IEEE Trans. Appl. Supercond.",
    volume = "35",
    number = "8",
    pages = "0601420",
    year = "2025"
}

@article{Ball:2022qks,
    author = "Ball, Richard D. and Candido, Alessandro and Cruz-Martinez, Juan and Forte, Stefano and Giani, Tommaso and Hekhorn, Felix and Kudashkin, Kirill and Magni, Giacomo and Rojo, Juan",
    collaboration = "NNPDF",
    title = "{Evidence for intrinsic charm quarks in the proton}",
    eprint = "2208.08372",
    archivePrefix = "arXiv",
    primaryClass = "hep-ph",
    reportNumber = "Nikhef 2021-032, Edinburgh 2021/28, TIF-UNIMI-2021-21",
    doi = "10.1038/s41586-022-04998-2",
    journal = "Nature",
    volume = "608",
    number = "7923",
    pages = "483--487",
    year = "2022"
}

@article{Kowitt:1993ns,
	Author = {Kowitt, M. S. and others},
	Collaboration = {E789},
	Doi = {10.1103/PhysRevLett.72.1318},
	Journal = {Phys. Rev. Lett.},
	Pages = {1318},
	Slaccitation = {%%CITATION = PRLTA,72,1318;%%},
	Title = {{Production of $J/\psi$ at large $x_F$ in 800~GeV/c pCu and pBe collisions}},
	Volume = {72},
	Year = {1994}}

@article{Johnson:2000dm,
	Archiveprefix = {arXiv},
	Author = {Johnson, M. B. and Kopeliovich, B. Z. and Tarasov, A. V.},
	Doi = {10.1103/PhysRevC.63.035203},
	Eprint = {hep-ph/0006326},
	Journal = {Phys. Rev.},
	Pages = {035203},
	Slaccitation = {%%CITATION = HEP-PH/0006326;%%},
	Title = {{Broadening of transverse momentum of partons propagating through a medium}},
	Volume = {C63},
	Year = {2001}}

@article{Vogt:2001ky,
	Author = {Vogt, R.},
	Eprint = {hep-ph/0107045},
	Journal = {Nucl. Phys.},
	Pages = {539},
	Slaccitation = {%%CITATION = HEP-PH 0107045;%%},
	Title = {{Are the $J/\psi$ and $\chi_c$ A-dependencies the same?}},
	Volume = {A700},
	Year = {2002}}

@article{Eskola:2016oht,
	Archiveprefix = {arXiv},
	Author = {Eskola, Kari J. and Paakkinen, Petja and Paukkunen, Hannu and Salgado, Carlos A.},
	Doi = {10.1140/epjc/s10052-017-4725-9},
	Eprint = {1612.05741},
	Journal = {Eur. Phys. J.},
	Pages = {163},
	Primaryclass = {hep-ph},
	Slaccitation = {%%CITATION = ARXIV:1612.05741;%%},
	Title = {{EPPS16: Nuclear parton distributions with LHC data}},
	Volume = {C77},
	Year = {2017}}

@misc{Vogt:2005ia,
	Author = {Vogt, R.},
	Eprint = {nucl-th/0507027},
	Slaccitation = {%%CITATION = NUCL-TH 0507027;%%},
	Title = {Baseline cold matter effects on $J/\psi$ production in A A collisions},
	Year = {2005}}

@article{NA38:1998lyg,
    author = "Abreu, M. C. and others",
    collaboration = "NA38",
    title = "{Charmonia production in 450-GeV/c proton induced reactions}",
    reportNumber = "CERN-EP-98-161",
    doi = "10.1016/S0370-2693(98)01398-7",
    journal = "Phys. Lett. B",
    volume = "444",
    pages = "516--522",
    year = "1998"
}

@article{NA60:2010wey,
    author = "Arnaldi, R and others",
    collaboration = "NA60",
    title = "{J/psi production in proton-nucleus collisions at 158 and 400 GeV}",
    eprint = "1004.5523",
    archivePrefix = "arXiv",
    primaryClass = "nucl-ex",
    doi = "10.1016/j.physletb.2011.11.042",
    journal = "Phys. Lett. B",
    volume = "706",
    pages = "263--267",
    year = "2012"
}

@article{Alde:1990wa,
    author = "Alde, D. M. and others",
    title = "{The A-Dependence of $J / \psi$ and $\psi^{\prime}$ Production at 800 GeV/c}",
    reportNumber = "FERMILAB-PUB-90-156-E",
    doi = "10.1103/PhysRevLett.66.133",
    journal = "Phys. Rev. Lett.",
    volume = "66",
    pages = "133--136",
    year = "1991"
}

@article{CMS:2022wfi,
    author = "Tumasyan, Armen and others",
    collaboration = "CMS",
    title = "{Nuclear modification of $\Upsilon$ states in pPb collisions at $\sqrt{s_\mathrm{NN}}$ = 5.02 TeV}",
    eprint = "2202.11807",
    archivePrefix = "arXiv",
    primaryClass = "hep-ex",
    reportNumber = "CMS-HIN-18-005, CERN-EP-2020-181",
    doi = "10.1016/j.physletb.2022.137397",
    journal = "Phys. Lett. B",
    volume = "835",
    pages = "137397",
    year = "2022"
}

@article{LHCb:2018jry,
    author = "Aaij, Roel and others",
    collaboration = "LHCb",
    title = "{First Measurement of Charm Production in its Fixed-Target Configuration at the LHC}",
    eprint = "1810.07907",
    archivePrefix = "arXiv",
    primaryClass = "hep-ex",
    reportNumber = "LHCb-PAPER-2018-023, CERN-EP-2018-266",
    doi = "10.1103/PhysRevLett.122.132002",
    journal = "Phys. Rev. Lett.",
    volume = "122",
    number = "13",
    pages = "132002",
    year = "2019"
}

@article{Koch:1986ud,
	Author = {Koch, P. and Muller, B. and Rafelski, J.},
	Journal = {Phys. Rept.},
	Pages = {167-262},
	Slaccitation = {%%CITATION = PRPLC,142,167;%%},
	Title = {STRANGENESS IN RELATIVISTIC HEAVY ION COLLISIONS},
	Volume = {142},
	Year = {1986}}

@article{AbdulKhalek:2021gbh,
    author = "Abdul Khalek, R. and others",
    title = "{Science Requirements and Detector Concepts for the Electron-Ion Collider}: {EIC Yellow Report}",
    eprint = "2103.05419",
    archivePrefix = "arXiv",
    primaryClass = "physics.ins-det",
    reportNumber = "BNL-220990-2021-FORE, JLAB-PHY-21-3198, LA-UR-21-20953",
    doi = "10.1016/j.nuclphysa.2022.122447",
    journal = "Nucl. Phys. A",
    volume = "1026",
    pages = "122447",
    year = "2022"
}

@article{NA50:2003fvu,
    author = "Alessandro, B. and others",
    collaboration = "NA50",
    title = "{Charmonium production and nuclear absorption in p-A interactions at 450~GeV}",
    reportNumber = "CERN-EP-2003-037",
    doi = "10.1140/epjc/s2003-01539-y",
    journal = "Eur. Phys. J. C",
    volume = "33",
    pages = "31--40",
    year = "2004"
}

@article{Anderson:1979tt,
    author = "Anderson, K. J. and others",
    title = "{Production of Muon Pairs by 225-GeV pi+-, K+, p+- Beams on Nuclear Targets}",
    reportNumber = "FERMILAB-PUB-79-104-E, EFI-79-6-CHICAGO",
    doi = "10.1103/PhysRevLett.42.944",
    journal = "Phys. Rev. Lett.",
    volume = "42",
    pages = "944",
    year = "1979"
}

@article{Leitch:1995yc,
    author = "Leitch, M. J. and others",
    title = "{Nuclear dependence of $J/\psi$ production by 800-GeV/c protons near $x_F$ = 0}",
    reportNumber = "FERMILAB-PUB-95-047",
    doi = "10.1103/PhysRevD.52.4251",
    journal = "Phys. Rev. D",
    volume = "52",
    pages = "4251--4253",
    year = "1995"
}

@article{HERA-B:2008ymp,
    author = "Abt, I. and others",
    collaboration = "HERA-B",
    title = "{Kinematic distributions and nuclear effects of J/$\psi$ production in 920-GeV fixed-target proton-nucleus collisions}",
    eprint = "0812.0734",
    archivePrefix = "arXiv",
    primaryClass = "hep-ex",
    reportNumber = "DESY-08-180",
    doi = "10.1140/epjc/s10052-009-0965-7",
    journal = "Eur. Phys. J. C",
    volume = "60",
    pages = "525--542",
    year = "2009"
}

@article{Eskola:2021nhw,
    author = "Eskola, Kari J. and Paakkinen, Petja and Paukkunen, Hannu and Salgado, Carlos A.",
    title = "{EPPS21: a global QCD analysis of nuclear PDFs}",
    eprint = "2112.12462",
    archivePrefix = "arXiv",
    primaryClass = "hep-ph",
    doi = "10.1140/epjc/s10052-022-10359-0",
    journal = "Eur. Phys. J. C",
    volume = "82",
    number = "5",
    pages = "413",
    year = "2022"
}

@article{PHENIX:2013pmn,
    author = "Adare, A. and others",
    collaboration = "PHENIX",
    title = "{Nuclear Modification of $\psi$, $\chi_c$, and $J/\psi$ Production in d+Au Collisions at $\sqrt{s_{NN}}$=200  GeV}",
    eprint = "1305.5516",
    archivePrefix = "arXiv",
    primaryClass = "nucl-ex",
    doi = "10.1103/PhysRevLett.111.202301",
    journal = "Phys. Rev. Lett.",
    volume = "111",
    number = "20",
    pages = "202301",
    year = "2013"
}

@article{PHENIX:2022nrm,
    author = "Acharya, U. A. and others",
    collaboration = "PHENIX",
    title = "{Measurement of $\psi(2S)$ nuclear modification at backward and forward rapidity in $p$ $+$ $p$, $p$ $+$ Al, and $p$ $+$ Au collisions at $\sqrt{s_{_{NN}}}=200$ GeV}",
    eprint = "2202.03863",
    archivePrefix = "arXiv",
    primaryClass = "nucl-ex",
    doi = "10.1103/PhysRevC.105.064912",
    journal = "Phys. Rev. C",
    volume = "105",
    number = "6",
    pages = "064912",
    year = "2022"
}

@article{PHENIX:2019brm,
    author = "Acharya, U. and others",
    collaboration = "PHENIX",
    title = "{Measurement of $J/\psi$ at forward and backward rapidity in $p+p$, $p+A$l, $p+A$u, and $^3$He$+$Au collisions at $\sqrt{s_{_{NN}}}=200~{\rm GeV}$}",
    eprint = "1910.14487",
    archivePrefix = "arXiv",
    primaryClass = "hep-ex",
    doi = "10.1103/PhysRevC.102.014902",
    journal = "Phys. Rev. C",
    volume = "102",
    number = "1",
    pages = "014902",
    year = "2020"
}

@article{Lourenco:2008sk,
    author = "Lourenco, Carlos and Vogt, Ramona and Woehri, Hermine K.",
    title = "{Energy dependence of J/psi absorption in proton-nucleus collisions}",
    eprint = "0901.3054",
    archivePrefix = "arXiv",
    primaryClass = "hep-ph",
    reportNumber = "CERN-PH-EP-2008-019",
    doi = "10.1088/1126-6708/2009/02/014",
    journal = "JHEP",
    volume = "02",
    pages = "014",
    year = "2009"
}

@article{HADES:2020ver,
    author = "Adamczewski-Musch, J. and others",
    collaboration = "HADES",
    title = "{Charged-pion production in $\mathbf {Au+Au}$ collisions at $\sqrt{\mathbf {s}_{\mathbf {NN}}} = 2.4~{\mathbf {GeV}}$: HADES Collaboration}",
    eprint = "2005.08774",
    archivePrefix = "arXiv",
    primaryClass = "nucl-ex",
    doi = "10.1140/epja/s10050-020-00237-2",
    journal = "Eur. Phys. J. A",
    volume = "56",
    number = "10",
    pages = "259",
    year = "2020"
}

@misc{NA60:2022sze,
    author = "Ahdida, C. and others",
    collaboration = "NA60+",
    title = "{Letter of Intent: the NA60+ experiment}",
    eprint = "2212.14452",
    archivePrefix = "arXiv",
    primaryClass = "nucl-ex",
    reportNumber = "CERN-SPSC-2022-036 / SPSC-I-259",
    month = "12",
    year = "2022"
}

@misc{STAR:2026kqj,
    collaboration = "STAR",
    title = "{Probing Late-Stage Hadronic Interactions at High Baryon Density via $K^{*0}$ Production in the RHIC Beam Energy Scan Program}",
    eprint = "2601.14884",
    archivePrefix = "arXiv",
    primaryClass = "nucl-ex",
    month = "1",
    year = "2026"
}

@article{Zhao:2010nk,
    author = "Zhao, Xingbo and Rapp, Ralf",
    title = "{Charmonium in Medium: From Correlators to Experiment}",
    eprint = "1008.5328",
    archivePrefix = "arXiv",
    primaryClass = "hep-ph",
    doi = "10.1103/PhysRevC.82.064905",
    journal = "Phys. Rev. C",
    volume = "82",
    pages = "064905",
    year = "2010"
}

\end{document}